\documentclass[aps,prd,notitlepage,showpacs,nofootinbib,superscriptaddress]{revtex4-1}
\usepackage{graphicx}
\usepackage[utf8]{inputenc} 
\usepackage{subfigure}
\usepackage{amsmath}
\usepackage{amssymb}
\usepackage{braket}
\usepackage{float}
\usepackage{slashed}
\usepackage[normalem]{ulem}
\usepackage{adjustbox}
\usepackage{multirow}
\usepackage{placeins}
\usepackage[normalem]{ulem} 
\usepackage[compat=1.1.0]{tikz-feynhand}
\usepackage{tikz-feynman}
\tikzfeynmanset{compat=1.1.0}
\usepackage{feynmp}
\usepackage{tikzsymbols}
\usetikzlibrary{shapes.misc,positioning,calc}
\usepackage{caption}
\usepackage{subcaption}
\usepackage{amsmath}
\usepackage{amssymb}
\usepackage{slashed}

\usepackage[usenames,dvipsnames]{xcolor}
\usepackage[colorinlistoftodos]{todonotes}
\usepackage[colorlinks=true,citecolor=darkred,urlcolor=darkred, pdfborder={0 0 0}]{hyperref}
\definecolor{darkred}{rgb}{0.6,0,0}
\definecolor{darkbrown}{rgb}{0.39,0.26,0.12}

\definecolor{linkcolor}{rgb}{0,0,0.5}
 
\newcommand {\ignore}[1]{}

\def\be{\begin{equation}}
\def\ee{\end{equation}}

\def\gsim{\raise0.3ex\hbox{$\;>$\kern-0.75em\raise-1.1ex\hbox{$\sim\;$}}}
\def\lsim{\raise0.3ex\hbox{$\;<$\kern-0.75em\raise-1.1ex\hbox{$\sim\;$}}}

\usepackage{soul}
\usepackage{changes}
\usepackage{comment}

\definecolor{mightnightblue}{RGB}{25,25,112}

\definecolor{brown}{rgb}{0.59, 0.29, 0.0}

\def\21{$\mathrm{SU(2)_L \otimes U(1)_Y}$}

\begin{document}
\begin{flushright}
CQUeST-2026-0800
\end{flushright}
\bibliographystyle{unsrt}   

\title{
\color{BrickRed} From Kination to Detection: Scotogenic Scalar Dark Matter} 

\author{Eung Jin Chun}\email{ejchun@kias.re.kr}
\affiliation{Korea Institute for Advanced Study, Seoul 02455, Korea}

\author{Ranjeet Kumar}
\email{kumarranjeet.drk@gmail.com}
\affiliation{Institute for Convergence of Basic Studies, Seoul National University of Science and Technology, Seoul 01811, Republic of Korea}

\author{Sanjoy Mandal}\email{smandal@kias.re.kr}
\affiliation{Korea Institute for Advanced Study, Seoul 02455, Korea}

\author{Abhishek Roy}\email{abhishek@sogang.ac.kr}
\affiliation{Center for Quantum Spacetime, Sogang University, 35 Baekbeom-ro, Mapo-gu, Seoul, 121-742, South Korea}
\affiliation{Department of Physics, Sogang University, 35 Baekbeom-ro, Mapo-gu, Seoul, 121-742, South Korea}

\begin{abstract}
\noindent
We investigate scalar dark matter in the scotogenic model during an early kination-dominated epoch. The enhanced expansion rate causes earlier freeze-out and increases the relic abundance, opening viable dark matter masses from approximately $200~\mathrm{GeV}$ to $1~\mathrm{TeV}$ while reducing the need for co-annihilation with a nearly degenerate singlet fermion. Current indirect searches strongly constrain the $(100-200)~\mathrm{GeV}$ region, whereas much of the surviving parameter space can be probed by CTA and next-generation direct detection experiments. The viable spectra also lead to complementary collider signatures, including missing transverse energy, displaced leptons, and disappearing charged tracks, providing promising tests at the HL-LHC and future lepton colliders.
\end{abstract}

\maketitle


\section{Introduction} \label{sec:intro}

Despite its remarkable success, the Standard Model (SM) cannot explain neutrino masses, the matter–antimatter asymmetry, or the particle nature of dark matter (DM), whose existence is strongly supported by astrophysical and cosmological observations \cite{Zwicky:1933gu,Rubin:1970zza,Rubin:1980zd,Planck:2018vyg}. Models connecting neutrino mass generation with a viable DM candidate are therefore particularly well motivated.
In this context, the scotogenic model \cite{Tao:1996vb,Ma:2006km} provides a minimal and elegant framework where neutrino masses are generated radiatively through interactions with a dark sector, while the same sector naturally contains a stable DM candidate.

The scotogenic model extends the SM by introducing an inert scalar doublet $\Phi$ and three gauge-singlet fermions $N_i$ ($i=1,2,3$). These new fields are odd under an exact $Z_2$ symmetry, while all SM fields are even. This symmetry forbids tree-level neutrino masses and stabilizes the lightest $Z_2$-odd particle, allowing either the lightest neutral scalar or the lightest singlet fermion to serve as a viable DM candidate. Neutrino masses are instead generated radiatively at the one-loop level through interactions involving the new scalar and fermionic fields. The model therefore establishes a direct connection between neutrino mass generation and the dark sector, while also giving rise to rich phenomenology involving DM relic abundance, direct and indirect detection, lepton flavor violation, and collider signatures.

The scotogenic model has consequently received considerable attention in the literature \cite{Fraser:2014yha,Fraser:2015mhb,Ma:2016mwh,Kang:2019sab,Mandal:2019oth,Avila:2019hhv,Borah:2021rbx,Batra:2022org,Batra:2022pej,CentellesChulia:2022vpz,Karan:2023adm,Chun:2023vbh,Kumar:2023moh,Kumar:2024zfb,CentellesChulia:2024iom,Lozano:2025tst,Roy:2025moo,Nomura:2025yoa,Kumar:2025cte,Avila:2025qsc,Kumar:2025zvv,Kang:2026osw,Kang:2026lgr}. Depending on the identity of the DM candidate and the dark sector mass spectrum, the observed relic abundance can be obtained through various annihilation and co-annihilation processes. Co-annihilation becomes particularly important when DM self-annihilation alone is insufficient, making the viable parameter space sensitive to the particle masses, couplings, and cosmological history.
Most previous studies assume a standard radiation-dominated Universe, in which the relic abundance is determined by conventional thermal freeze-out \cite{Scherrer:1985zt}. However, the pre-Big Bang nucleosynthesis (BBN) expansion history is not directly constrained to be radiation dominated. A non-standard cosmological background can modify the expansion rate and, consequently, the freeze-out dynamics and predicted relic abundance. This motivates revisiting the DM phenomenology and viable parameter space of the scotogenic model beyond the standard cosmological scenario.
\par Motivated by this, in the present work we examine the phenomenology of scalar DM in the scotogenic model within a non-standard cosmological scenario characterized by an early period of kination domination \cite{Zeldovich:1961sbr,Spokoiny:1993kt,Joyce:1996cp,Ferreira:1997hj}. In this scenario, the energy density of the Universe is dominated by the kinetic energy of a scalar field, which redshifts faster than radiation and leads to an enhanced expansion rate compared to the standard radiation-dominated case. Consequently, the DM freeze-out dynamics and the resulting relic abundance can be significantly modified.
In particular, the faster expansion rate can cause DM to decouple earlier from the thermal bath, thereby enhancing its final relic abundance. This opens up the possibility of rendering phenomenologically viable regions of the scotogenic parameter space that would otherwise predict an underabundant DM density in the standard cosmological scenario. We therefore investigate the impact of a kination-dominated epoch on the DM relic abundance and explore its implications for the viable parameter space and DM phenomenology of the model.

We first consider the standard cosmological scenario to identify the underabundant DM parameter space that can be rendered viable by kination. 
We focus on the case where only $N_1$ participates in co-annihilation. In this case, the observed relic abundance can be obtained for $M_{\rm DM}\gtrsim300$ GeV, whereas lighter DM remains underabundant in the standard cosmology.
We then investigate whether kination can raise the relic abundance of these underabundant points to the observed $2\sigma$ range \cite{Planck:2018vyg}. For each point in the mass range $100~\mathrm{GeV}\lesssim M_{\rm DM}\lesssim1~\mathrm{TeV}$, we determine the reheating temperature $T_{\rm RH}$ required to reproduce the observed abundance. Although the region $100~\mathrm{GeV}\lesssim M_{\rm DM}\lesssim200$ GeV is excluded by current indirect detection constraints \cite{Alvarez:2020cmw,Slatyer:2015jla}, the surviving region up to $1$ TeV lies within the projected sensitivity of CTA \cite{CTA:2020qlo}. These underabundant points typically have $\langle\sigma v\rangle_{\rm eff}\gtrsim3\times10^{-26} \mathrm{cm^3s^{-1}}$, making them relevant for indirect detection and complementary direct detection searches. The impact of kination is most pronounced for $M_{\rm DM}\lesssim300$ GeV, where a larger enhancement of the relic abundance is generally required. At higher masses, the observed abundance can be obtained over a broader range of $T_{\rm RH}$ and mass splittings $M_{N_1}-M_{A^0}$, often without significant $N_1$ co-annihilation.

Finally, we investigate the collider phenomenology of the cosmologically viable parameter space at the HL-LHC and future lepton colliders. We first consider the inert doublet model (IHDM) as a reference and then examine the additional signatures arising in the scotogenic model. In particular, the compressed dark sector spectrum can give rise to long-lived charged scalars and singlet fermions, leading to disappearing tracks and displaced vertices, while larger mass splittings and Yukawa couplings can yield prompt leptons accompanied by missing transverse energy. We discuss how kination opens up additional low-mass regions with potentially larger electroweak production rates, offering complementary collider signatures of scalar DM.

\section{The Model}
\label{sec:model}

We begin with a brief overview of the scotogenic model, which provides a unified framework for simultaneously addressing the origin of neutrino masses and the stability of the DM candidate. The model extends the SM particle content by introducing an inert scalar $\Phi$ and three fermions $N_i$ ($i=1,2,3$). The scalar $\Phi$ transforms as an $SU(2)_L$ doublet with hypercharge $Y=1$, while the fermions $N_i$ are singlets under the SM gauge group. In addition, the model is supplemented by a discrete $Z_2$ symmetry under which the newly introduced particles are odd, whereas all SM fields remain even. This symmetry ensures the stability of the lightest $ Z_2$-odd particle, thereby providing a viable DM candidate. The particle content and their transformation under various symmetries are summarized in Table \ref{tab:fields}.
 \begin{figure}[!t]
    \centering
    \input {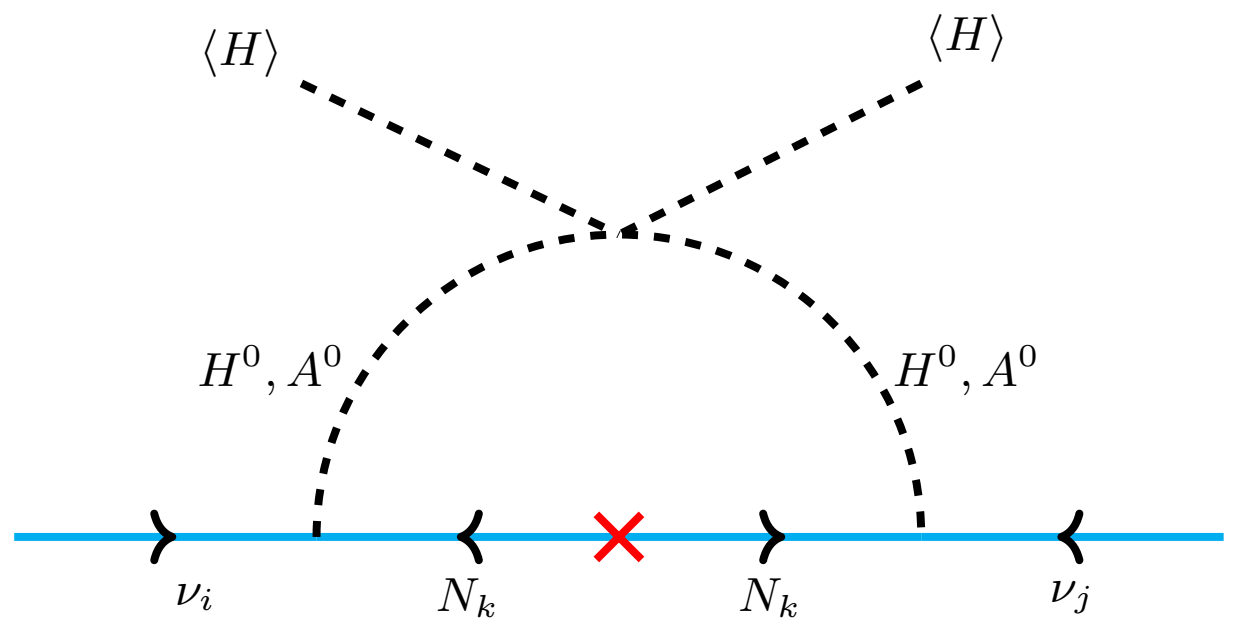}
    \caption{\centering Feynman diagram for one-loop neutrino mass generation in the scotogenic model.}
    \label{fig:feyn_numass}
\end{figure}
\begin{table}[!h]
    \centering
    \begin{tabular}{|c|c||c|c|c|}
        \hline 
     & \  Fields \ & \  $SU(2)_L$ \ & \ $U(1)_Y$ \ & $Z_2$ \ \\ \hline \hline
        \ \multirow{3}{*}{\rotatebox{90}{SM}} \ & $L_i$ &  $2$ & $-1$ & $+$  \\
      &  $\ell_{R_i}$ & $1$ & $-2$  & $+$ \\ 
     &   $H$ & $2$ & $~~1$ & $+$  \\
        \hline \hline
       \ \multirow{2}{*}{\rotatebox{90}{BSM}} \ 
    &     $\Phi$ & $2$ & $~~1$ & $-$ \\
     &   $N_i$ & $1$ & $~~0$ & $-$ \\
        \hline
    \end{tabular}
    \caption{Particle content and their transformation properties under different symmetries, where $i=1,2,3$. The discrete $Z_2$ symmetry serves as a dark symmetry to protect the stability of the DM candidate.}
    \label{tab:fields}
\end{table}
%

Following the charge assignments of the fields provided in Table \ref{tab:fields}, the renormalizable gauge-invariant terms in the Lagrangian that generate neutrino mass are given by,
\begin{equation}
-\mathcal{L}_{\text{Y}} \supset Y^{\ell}_{\alpha i} \overline{L}_{\alpha} H \ell_{R_i} + Y^{\nu}_{i\alpha } \,  \overline{N}_i \tilde{\Phi}^\dagger L_{\alpha} + \frac{1}{2} M_{N_k} \overline{N_k^c}N_k +\text{h.c.},
\label{eq:LagYL}
\end{equation}
where $\tilde{\Phi}=i\sigma_2\Phi^*$ and $L=(\nu_{\ell L},\ell_L)^T$. We take the fermion mass matrix $M_{N_k}$ to be diagonal. 
 As a consequence of the imposed $Z_2$ symmetry, the conventional tree-level seesaw mechanism is absent. Neutrino masses are instead generated radiatively through the one-loop diagram involving the singlet fermions and inert scalars as shown in Fig. \ref{fig:feyn_numass}. 
 The corresponding neutrino mass matrix is given by \cite{Ma:2006km,Mandal:2021yph,DeRomeri:2022cem,Chun:2023vbh}:
\begin{align}
\label{eq:numass}
(m_\nu)_{ij} &
= \sum_{k=1}^3 \frac{Y^\nu_{ik} \, Y^{\nu T}_{kj} M_{N_k}}{32 \pi^2}\left[ \frac{M_{H^0}^2}{M_{H^0}^2-M_{N_k}^2} \log \frac{M_{H^0}^2}{M_{N_k}^2} -  \frac{M_{A^0}^2}{M_{A^0}^2 - M_{N_k}^2} \log \frac{M_{A^0}^2}{M_{N_k}^2} \right],\\
& \equiv (Y^{\nu T} \Lambda Y^{\nu})_{ij},
\end{align}
where $\Lambda$ denotes the diagonal matrix $\Lambda=\text{diag}(\Lambda_1,\Lambda_2,\Lambda_3)$, with
\begin{align}
\label{eq:Lambda}
\Lambda_k
=  \frac{M_{N_k}}{32 \pi^2}\left[ \frac{M_{H^0}^2}{M_{H^0}^2-M_{N_k}^2} \log \frac{M_{H^0}^2}{M_{N_k}^2} -  \frac{M_{A^0}^2}{M_{A^0}^2 - M_{N_k}^2} \log \frac{M_{A^0}^2}{M_{N_k}^2} \right].
\end{align}
To relate the model parameters to the experimentally measured neutrino masses and mixing parameters, we make use of the Casas-Ibarra parametrization~\cite{Casas:2001sr}, according to which the Yukawa coupling matrix takes the following form:
\begin{align}
Y^\nu = \sqrt{\Lambda}^{-1} R \sqrt{\hat{m}_\nu} U^{\dagger}_{\rm lep}.
\label{eq:Ynu}
\end{align}
Here, $\hat{m}_\nu$ corresponds to the diagonal matrix of light neutrino masses and $U^{\dagger}_{\rm lep}$ is the Pontecorvo-Maki-Nakagawa-Sakata (PMNS) matrix responsible for diagonalizing the neutrino mass matrix. $R$ is a general $3 \times 3$ orthogonal matrix which can be expressed as the product of three rotation matrices,
\begin{align} \label{eq:Rmat}
R = R_{23}(x).R_{13}(y).R_{12}(z),
\end{align}
where $x$, $y$, and $z$ denote the rotation angles in the (2,3), (1,3), and (1,2) planes, respectively. In the present analysis, we restrict ourselves to the simplest realization by choosing $R$ to be the identity matrix:
$
R=\mathbb{I},
$
which corresponds to setting ($x=y=z=0$). This benchmark choice eliminates the additional degrees of freedom associated with the Casas-Ibarra parametrization, allowing us to investigate whether the model can simultaneously reproduce the observed neutrino data while satisfying the DM relic abundance, direct and indirect detection limits, and charged lepton flavor violating constraints within a minimal parameter space. We would like to stress that this assumption is adopted solely to simplify the analysis and does not represent the most general scenario. Relaxing this condition by allowing non-zero (real and/or complex) rotation angles can modify the Yukawa structure, potentially enlarging the viable parameter space and altering the predictions for flavor observables, particularly charged lepton flavor violation. 
Throughout this work, we assume the normal ordering of neutrino masses and vary the lightest neutrino mass in the range $m_{\nu_1} \sim(10^{-5}-0.03)$ eV. The neutrino oscillation parameters are fixed to their current best-fit values from the NuFIT analysis \cite{Esteban:2024eli,Esteban:2026phq}.

\FloatBarrier

The scalar sector consists of the SM Higgs doublet $H$ and an inert scalar doublet $\Phi$. With the charge assignments given in Table~\ref{tab:fields}, the most general renormalizable scalar potential consistent with the gauge and discrete symmetries is given by
\begin{align}
V =  &\mu_1^2 H^\dagger H + \mu_2^2 \Phi^\dagger \Phi + \frac{\lambda_1}{2} (H^\dagger H)^2 + \frac{\lambda_2}{2} (\Phi^\dagger \Phi)^2 + \lambda_3 (H^\dagger H)(\Phi^\dagger \Phi) + \lambda_4 (\Phi^\dagger H) (H^\dagger \Phi )+ \nonumber \\  &\frac{\lambda_5}{2} \left[ (\Phi^\dagger H)^2 + \text{H.c.} \right].\label{eq:potential}
\end{align}
%
Adopting the unitary gauge, the SM Higgs and inert scalar doublets are parametrized as,
\begin{align}
H=
\begin{pmatrix}
0\\
(v+h)/\sqrt{2}
\end{pmatrix}, \quad 
\Phi=
\begin{pmatrix}
H^+\\
(H^0+iA^0)/\sqrt{2}
\end{pmatrix},
\label{eqmass}
\end{align}
where $v= 246$ GeV is the SM Higgs vacuum expectation value (VEV), and $h$ is the SM Higgs boson with mass $M_h\simeq 125\,\rm{GeV}$. The masses of new scalar particles can be expressed as, 
\begin{align}
M_{H^{+}}^{2}=\mu_{2}^{2}+\frac{1}{2}\lambda_{3}v^{2},\,\,
M_{H^{0}}^{2}=\mu_{2}^{2}+\lambda_{L}v^{2},\,\,
M_{A^{0}}^{2}=\mu_{2}^{2}+\overline{\lambda}_{L}v^{2},
\end{align}
where we have defined
\begin{align}
\lambda_{L}=\frac{1}{2}\left(\lambda_{3}+\lambda_{4}+\lambda_{5}\right)\,\,\,\text{and} \,\,\, \overline{\lambda}_{L}= \frac{1}{2}\left(\lambda_{3}+\lambda_{4}-\lambda_{5}\right).
\end{align}
We consider $\lambda_5 >0$, which leads to $\overline{\lambda}_{L} < \lambda_{L}$ and $M_{A^{0}} < M_{H^{0}}$. In addition, we require $M_{A^0} < M_{H^+}, \ M_{N_k}$, ensuring that $A^0$ is the lightest $Z_2$ odd particle and therefore serves as a DM candidate. 

The scalar potential in Eq.~\ref{eq:potential} introduces seven independent parameters in the scalar sector, consisting of two mass parameters, $\mu^2_1$, $\mu^2_2$, along with five dimensionless quartic couplings $\lambda_i$. Once the $H$ acquires VEV, $\mu_1^2$ is related with $\lambda_1$ and can be determine by the observed Higgs mass: $M_h\simeq 125\,\rm{GeV}$. 
The coupling $\lambda_2$ does not affect the tree-level scalar mass spectrum or the dark sector phenomenology. Therefore, we fix $\lambda_2 =0.1$ throughout our analysis, which satisfies the relevant theoretical constraints as discussed next. 
The remaining parameters $\mu^2_2$ and $\lambda_{3,4,5}$ determine the masses of the dark sector scalars. In addition, the couplings $\lambda_3$ and $\lambda_4$ are crucial for the DM phenomenology, especially regarding the constraints from direct detection experiments. 
Furthermore, the coupling $\lambda_5$ plays an important role in generating neutrino masses. When $\lambda_5 \to 0$, the mass splitting between the CP-even $H^0$ and CP-odd $A^0$ components of the inert doublet disappears, resulting in an exact cancellation between their contributions to the one-loop neutrino mass diagram. Consequently, the derived neutrino mass formula in Eq.~\ref{eq:numass} vanishes in this limit.
The scanned ranges of these parameters, along with the masses of the right handed neutrinos, are listed in Table~\ref{tab:range}. 
\begin{table}[]
\begin{center}
\begin{tabular}{| c | c | c | c |}
  \hline 
  Parameter    &   Range   &   Parameter    &   Range (in GeV)  \\
\hline
$\lambda_{2}$     &  	 $0.1$            &
$\mu_2$   &  $[10,10^4]$            \\
$\lambda_{3}$   &   $\pm[10^{-8},1]$           &
$M_{N_1}$   &   $M_{A^0}+[10^{-3},10^4]$   	   \\
$\lambda_{4}$   &   $\pm[10^{-8},1]$   	   &
$M_{N_2}$   &   $[10^{4},10^7]$   	   \\
$\lambda_{5}$          &	     $[10^{-10},1]$   &
$M_{N_3}$    &	  $[10^{4},10^7]$       \\
    \hline
  \end{tabular}
\end{center}
\caption{Input parameter ranges adopted in the numerical scan. Throughout the analysis, the inert scalar $A^0$ is considered to be the DM candidate.}
 \label{tab:range} 
\end{table}

\section{Constraints} \label{sec:const}

To explore the parameter space of the model, we perform a systematic numerical scan over the ranges specified in Table \ref{tab:range}. At each scan point, we require the model to satisfy a comprehensive set of theoretical consistency conditions, including vacuum stability, perturbativity, perturbative unitarity, and the realization of the inert vacuum as the global minimum, along with the relevant experimental constraints. The parameter points that survive these requirements are then employed to study the DM phenomenology, including the relic abundance, direct and indirect detection constraints, and the prospects for future experimental searches.
In the following, we first summarize the theoretical consistency conditions imposed on the model and subsequently list the experimental constraints used to identify the phenomenologically viable parameter space.

\subsection{Theoretical Constraints}

To ensure the theoretical consistency of the model, we impose a set of theoretical constraints on the scalar sector throughout our analysis. The corresponding conditions adopted in this work are summarized below:
\begin{align} \label{eq:vac}
    \lambda_1 > 0, \quad \lambda_2 >0, \quad \lambda_3 +\sqrt{\lambda_1 \lambda_2}>0, \quad \lambda_3 + \lambda_4 - |\lambda_5| + \sqrt{\lambda_1 \lambda_2}>0, \ \text{and} \  \lambda_4 - |\lambda_5|<0.
\end{align}
\begin{align} \label{eq:pert}
    &\lambda_2 < \frac{2 \pi}{3}, \quad \lambda_3 < 4 \pi, \quad \lambda_4 < 4 \pi, \quad |\lambda_5| < 4\pi, \quad |\lambda_3 + \lambda_4| < 4 \pi, \nonumber \\ 
    & |\lambda_4 \pm \lambda_5| < 8 \pi, \quad |\lambda_3 + \lambda_4 \pm \lambda_5| < 4 \pi. 
\end{align}
\begin{align} \label{eq:pertUni}
    & |\lambda_3 \pm \lambda_4| < 8 \pi, \quad |\lambda_3 \pm \lambda_5| < 8 \pi, \quad |\lambda_3 + 2\lambda_4 \pm 3 \lambda_5| < 8 \pi, \nonumber \\ 
    & \lambda_1 + \lambda_2 \pm \sqrt{(\lambda_1-\lambda_2)^2 + 4 \lambda_4^2}| < 16 \pi, \quad  |\lambda_1 + \lambda_2 \pm \sqrt{(\lambda_1-\lambda_2)^2 + 4 \lambda_5^2}| < 16 \pi, \nonumber \\
    &|3\lambda_1 + 3\lambda_2 \pm \sqrt{9(\lambda_1-\lambda_2)^2 + 4 (2\lambda_3 + \lambda_4)^2}| < 16 \pi, \quad
\end{align}
Among these conditions, Eq.~\ref{eq:vac} ensures that the scalar potential is bounded from below and avoids the charge-breaking vacua \cite{Belyaev:2016lok,Ginzburg:2010wa}.  Eq.~\ref{eq:pert} guarantees that the quartic couplings remain within the perturbative regime and that one-loop corrections do not dominate over the corresponding tree-level contributions \cite{Garcia-Cely:2015khw,Belanger:2022qxt}. The conditions in Eq.~\ref{eq:pertUni} impose the perturbative unitarity constraints on all tree-level $2 \to 2$ scalar scattering amplitudes \cite{Arhrib:2012ia}.

\subsection{Experimental Constraints}

We further restrict the theoretically viable parameter space by imposing the current experimental constraints relevant to the model. These constraints are as follows: 
\begin{itemize}
    \item \textbf{LEP constraints:} The LEP measurements provide important constraints on the scalar sector of the model \cite{LEP:2000pgt}. In particular, they exclude SM gauge boson decays into dark sector particles, while the searches for singly charged scalar production, $e^+ e^- \to H^+ H^-$, further impose lower bounds on the inert scalar masses. The corresponding constraints are given by:
    \begin{align}
        M_{H^0} + M_{A^0} > M_Z, \quad 2 M_{H^+} > M_{Z}, \quad  M_{H^+} + M_{H^0} (M_{A^0}) > M_W, \quad \text{and} \quad M_{H^+} > 70 \ \text{GeV}.
    \end{align}
    \item \textbf{Higgs invisible decay:} For sufficiently light inert scalars, the Higgs boson can decay invisibly into dark sector particles via $h \to A^0 A^0$ and/or $h \to H^0 H^0$. The experimentally allowed upper limit on the invisible Higgs branching ratio is given by \cite{ParticleDataGroup:2024cfk}:
    \begin{align}
        \text{BR}(h \to \text{inv}) < 0.107.
    \end{align}
    \item \textbf{Higgs diphoton rate:} The signal strength for the Higgs boson decaying into two photons is defined as
    \begin{align}
        R_{\gamma \gamma} = \frac{\Gamma (h \to \gamma \gamma)}{\Gamma^{\rm SM} (h \to \gamma \gamma)}.
    \end{align}
    The current measurements of the Higgs diphoton signal strength $R_{\gamma \gamma}$ reported by the ATLAS \cite{ATLAS:2022tnm} and CMS \cite{CMS:2021kom} collaborations are given by
    \begin{align}
        R_{\rm ATLAS} =  1.04^{+0.10}_{-0.09}, \quad  \quad  R_{\rm CMS} = 1.12\pm 0.09. 
    \end{align}
    We combine the ATLAS and CMS measurements by constructing a  $\chi^2$ function following Refs. \cite{Kraml:2019sis,Belanger:2024wca}, given by
    \begin{align}
        \chi^2 = \sum_i \frac{(\mu_i-\hat{\mu})^2}{\sigma^+ \sigma^- + (\sigma^+ - \sigma^-)(\mu_i - \hat{\mu})},
    \end{align}
    here, $\mu_i$ and $\sigma^{\pm}$ denote the experimentally measured central values and their corresponding uncertainties, while $\hat{\mu}$ represents the model prediction. Throughout our analysis, we retain only those parameter points satisfying $\Delta \chi^2= \chi^2-\chi^2_{\rm min} < 4$, corresponding to the 95\% confidence level. 
     \item \textbf{Electroweak precision data:} New physics contributions to electroweak observables are conveniently parametrized in terms of the oblique parameters $S$, $T$, and $U$, which quantify deviations from the SM predictions through corrections to the electroweak gauge boson self-energies. For our analysis, the $S$ and $T$ parameters are calculated using the formalism presented in Ref. \cite{Belyaev:2016lok,Barbieri:2006dq}. The latest global electroweak fit \cite{Lu:2022bgw} yields the following values for the oblique parameters S and T:
     \begin{align}
         S_{\rm exp} = 0.05 \pm 0.08, \quad T_{\rm exp} = 0.09 \pm 0.07,
     \end{align}
     assuming $U=0$, and a correlation coefficient $\rho_{ ST}=+0.92$, and SM Higss mass $M_{h}=125$ GeV. We define the $\chi^2$ as
     \begin{equation}
\chi^2 = \frac{1}{\left(1-\rho_{ST}^{2}\right)}
\left[\frac{(S-S_{\rm exp})^2}{(\Delta S_{\rm exp})^2}+\frac{(T-T_{\rm exp})^2}{(\Delta T_{\rm exp})^2}-\frac{2\rho_{ST}(S-S_{\rm exp})(T-T_{\rm exp})}{\Delta S_{\rm exp}\,\Delta T_{\rm exp}}\right].
\label{eq:chi2_ST}
\end{equation}
where $\Delta S_{\rm exp}$ and $\Delta T_{\rm exp}$ denotes the $1\sigma$ experimental uncertainties. Throughout our analysis, we retain only those parameter points satisfying $\Delta \chi^2= \chi^2-\chi^2_{\rm min} < 4$, which corresponds to the 95\% confidence level. 
    \item \textbf{Charged lepton flavor violation:} In our analysis, we also consider the limit from charged lepton flavor violation (cLFV). These cLFV processes are induced by the same Yukawa interaction $Y^{\nu}_{i \alpha} \overline{N}_i \tilde{\Phi}^\dagger {L}_{\alpha} $ that generates the neutrino masses. 
    The two constraining processes are the two body $\ell_{\alpha} \to \ell_{\beta} \gamma$ and three body decay $\ell_{\alpha} \to 3 \ell_{\beta}$. For the three body decay, the most constraining one arises from $\mu \to 3 e$. The current experimental limits on these processes are as follows \cite{Belle:2021ysv,MEGII:2023ltw,BaBar:2009hkt}:
    \begin{align}
        &\text{Br}(\mu \to e \gamma) \leq 3.1 \times 10^{-13}, \\
        &\text{Br}(\tau \to e \gamma) \leq 3.3 \times 10^{-8}, \\
        &\text{Br}(\tau \to \mu \gamma) \leq 4.2 \times 10^{-8}, \\
        &\text{Br}(\mu \to 3e ) \leq 1.2 \times 10^{-12}.
    \end{align}
\end{itemize}

\section{DM Production in Radiation and Kination Eras}
We now discuss the DM relic abundance and the corresponding viable mass range in this model. The observed relic abundance is determined by the thermal evolution of DM in the early Universe. The interplay between annihilation and cosmological expansion determines when DM freezes out and, consequently, its present-day relic density. The thermal evolution of the DM number density $n_{A^0}$ is governed by the Boltzmann equation~\cite{Kolb:1990vq},
\begin{equation} \label{eq:Beq_A0}
\frac{dn_{A^0}}{dt}+3Hn_{A^0}=-\langle\sigma v\rangle_{A^0 A^0}\left(n_{A^0}^2-(n_{A^0}^{\rm eq})^2\right),
\end{equation}
where $n_{A^0}$ and $n_{A^0}^{\rm eq}$ are the actual and equilibrium number densities of the DM candidate, respectively. Here, $H$ is the Hubble expansion rate and $\langle\sigma v\rangle_{A^0 A^0}$ is the thermally averaged annihilation cross-section of $A^0$. This equation describes the evolution of the DM abundance when $A^0$ annihilations dominate, and co-annihilation effects can be neglected. When a $Z_2$-odd state $i$ has a mass splitting $(M_i-M_{A^0})$ that is not much larger than the DM freeze-out temperature $T_f \sim M_{A^0}/(20\text{--}30)$, its equilibrium abundance is not strongly Boltzmann suppressed relative to that of the DM, and can be expressed approximately as~\cite{Griest:1990kh},
\begin{align}
\frac{n_i^{\rm eq}}{n_{A^0}^{\rm eq}}
\simeq
\frac{g_i}{g_{A^0}}
\left(\frac{M_i}{M_{A^0}}\right)^{3/2}
e^{-(M_i-M_{A^0})/T_f},
\end{align}
where $g_i$ and $g_{A^0}$ denote the internal degrees of freedom of species $i$ and $A^0$, respectively. Such states can therefore contribute to the total annihilation rate, and their annihilation and co-annihilation channels must be included when determining $A^0$ evolution in the early Universe. If the co-annihilating species remain in chemical equilibrium with the DM until freeze-out, a single effective Boltzmann equation describes the coupled evolution of the total dark sector number density. The corresponding equation can be written as,
\begin{equation} \label{eq:noden}
\frac{dn}{dt}+3Hn=-\langle\sigma v\rangle_{\rm eff} \left(n^2-(n^{\rm eq})^2\right),
\end{equation}
where $n=\sum_{i=1}^N n_i$ and $n^{\rm eq}=\sum_{i=1}^N n_i^{\rm eq}$ are the total actual and equilibrium number densities of the dark sector, respectively, and $N$ corresponds to the number of dark sector particles. The effective annihilation cross-section $\langle\sigma v\rangle_{\rm eff}$ accounts for all relevant annihilation and co-annihilation processes and can be written as~\cite{Griest:1990kh,Profumo:2006bx},
\begin{equation}\label{Eqn:sigeff}
\langle \sigma v\rangle_{\rm eff}=\sum_{i,j=1}^N\ \langle \sigma v\rangle_{ij}\frac{g_ig_j}{(g_i\left(1+\Delta_i\right)^{3/2}{\rm e}^{-x\Delta_i})^2}\left(1+\Delta_i\right)^{3/2}\left(1+\Delta_j\right)^{3/2}{\rm e}^{-x(\Delta_i+\Delta_j)},
\end{equation}
where $\Delta_i=(M_i-M_{A^0})/M_{A^0}$ is the relative  mass splitting between the $i$-th dark sector particle and $A^0$, while $x=M_{A^0}/T$. The quantity $\langle \sigma v\rangle_{ij}$ denotes the thermally averaged cross-section for annihilation involving species $i$ and $j$. Equation~\ref{Eqn:sigeff} shows that additional co-annihilating states do not necessarily enhance the effective annihilation cross-section. Their contribution can either increase or decrease $\langle\sigma v\rangle_{\rm eff}$, depending on how the annihilation and co-annihilation rates involving these states compare with the annihilation cross-section of $A^0$. The validity of the effective Boltzmann equation in Eq. \ref{Eqn:sigeff} requires the co-annihilating species to remain in chemical equilibrium with the DM until freeze-out. In this case, their relative abundances track the corresponding equilibrium values, $r_i \approx r_i^{\rm eq}$, where $r_i=n_i/n$ and $r_i^{\rm eq}=n_i^{\rm eq}/n^{\rm eq}$. This condition is maintained by sufficiently rapid conversion processes between the co-annihilating species and the DM. Note that in the limit when the mass splitting between the $Z_2$-odd state $i$ and $A^0$ becomes larger than $T_f$, $\langle\sigma v\rangle_{\rm eff}$ tends to $\langle\sigma v\rangle_{A^0A^0}$ and in principle the Boltzmann equation with $\langle \sigma v\rangle_{\rm eff}$ mentioned in Eq.~\ref{eq:noden} becomes the one-species Boltzmann equation for $A^0$ given in Eq.~\ref{eq:Beq_A0}.

In the present model, the inert scalars ($H^\pm,H^0$) remain in chemical equilibrium with $A^0$ through electroweak gauge interactions.\footnote{We assume $N_{2,3}$ are much heavier than $A^0$; their equilibrium abundances are Boltzmann suppressed during freeze-out; therefore, neglect the contributions of $N_{2,3}$ to the evolution of the $A^0$ number density.}
The validity of the effective cross-section formalism therefore depends, in particular, on whether the conversion processes $N_1 \leftrightarrow A^0 \nu$ are efficient enough to maintain chemical equilibrium between $N_1$ and $A^0$ before freeze-out. We therefore adopt the following approximate criterion for maintaining chemical equilibrium between $N_1$ and $A^0$~\cite{Bandyopadhyay:2011qm, Frumkin:2021zng,Frumkin:2025dxq,Belanger:2025wjh}:
\begin{align} \label{eq:decaywidth_validity}
\Gamma_{N_1\to A^0\nu} \gtrsim \frac{M_{N_1}-M_{A^0}}{M_{A^0}} H(x=1),
\end{align}
where $\Gamma_{N_1\to A^0\nu}$ is the decay width of $N_1$ to $A^0$, which is given by,
\begin{align} \label{eq:decayexp}
\Gamma_{N_1\to A^0\nu}=\frac{ ({Y^{\nu}_{1\rm{eff}}})^2 M_{N_1}}{32 \pi}\left( 1-\frac{M^2_{A^0}}{M^2_{N_1}}\right)^2.
\end{align}
where $Y^{\nu}_{1\rm{eff}} \equiv \sqrt{\sum_\alpha(Y^{\nu}_{1\alpha})^2}$. In our analysis, we use the condition in Eq.~\ref{eq:decaywidth_validity} to identify the parameter space where conversion between $N_1$ and $A^0$ is sufficiently rapid to maintain chemical equilibrium, as required for solving the Boltzmann equation using the effective annihilation cross-section formalism. 

As the Universe expands and cools, the interaction rate of DM eventually becomes smaller than the expansion rate, leading to thermal freeze-out and leaving behind a relic abundance that depends on the particle masses, couplings, and the cosmological expansion history. In our numerical analysis, we compute the relic density using \texttt{micrOMEGAs7.1}~\cite{Belanger:2026asz}, which numerically solves the Boltzmann equation in Eq.~\ref{eq:noden} and incorporates all relevant annihilation and co-annihilation channels in the model.
It is important to emphasize that DM evolution before BBN depends sensitively on the expansion history of the Universe. The standard thermal freeze-out scenario assumes an instantaneous inflaton-decay approximation under which the reheating temperature is identified by the maximum temperature attained. Consequently, the Universe is assumed to remain radiation-dominated until the epoch of matter-radiation equality. However, reheating can be non-instantaneous, and the Universe may undergo a non-standard expansion history before BBN, potentially altering DM production and freeze-out dynamics.
 To investigate the impact of the cosmological background on the DM phenomenology, we study DM production in both the conventional radiation-dominated cosmology and a non-standard kination cosmology. In the kination scenario, the energy density of the Universe is dominated by the kinetic energy of a scalar field before the onset of radiation domination. Comparing these scenarios enables us to assess the impact of the cosmological background on the viable parameter space and the DM phenomenology of the scotogenic model.

\subsection{Radiation Dominated Universe}

We begin with the standard radiation-dominated cosmology, which serves as the reference scenario for our analysis. 
In this cosmological scenario, the Hubble expansion rate $H$ is expressed as 
\begin{align} \label{eq:rad_H}
    H^2= \frac{\rho_R}{3 M^2_P},
\end{align}
here $M_P$ is the reduced Planck mass, $M_P \simeq 2.4 \times 10^{18}$ GeV and $\rho_R$ is the SM energy density which is given by
\begin{align}
    \rho_R(T) = \frac{\pi^2}{30} g_{\rm eff} (T) T^4,
\end{align}
where $g_{\rm eff}(T)$ is the effective number of degrees of freedom which contribute to the SM energy density and $T$ is the temperature of SM photons. 
Accordingly, the DM relic abundance is obtained through the standard thermal freeze-out mechanism in the radiation-dominated Universe.  

\begin{figure} []
    \centering
\includegraphics[width=0.55\linewidth]{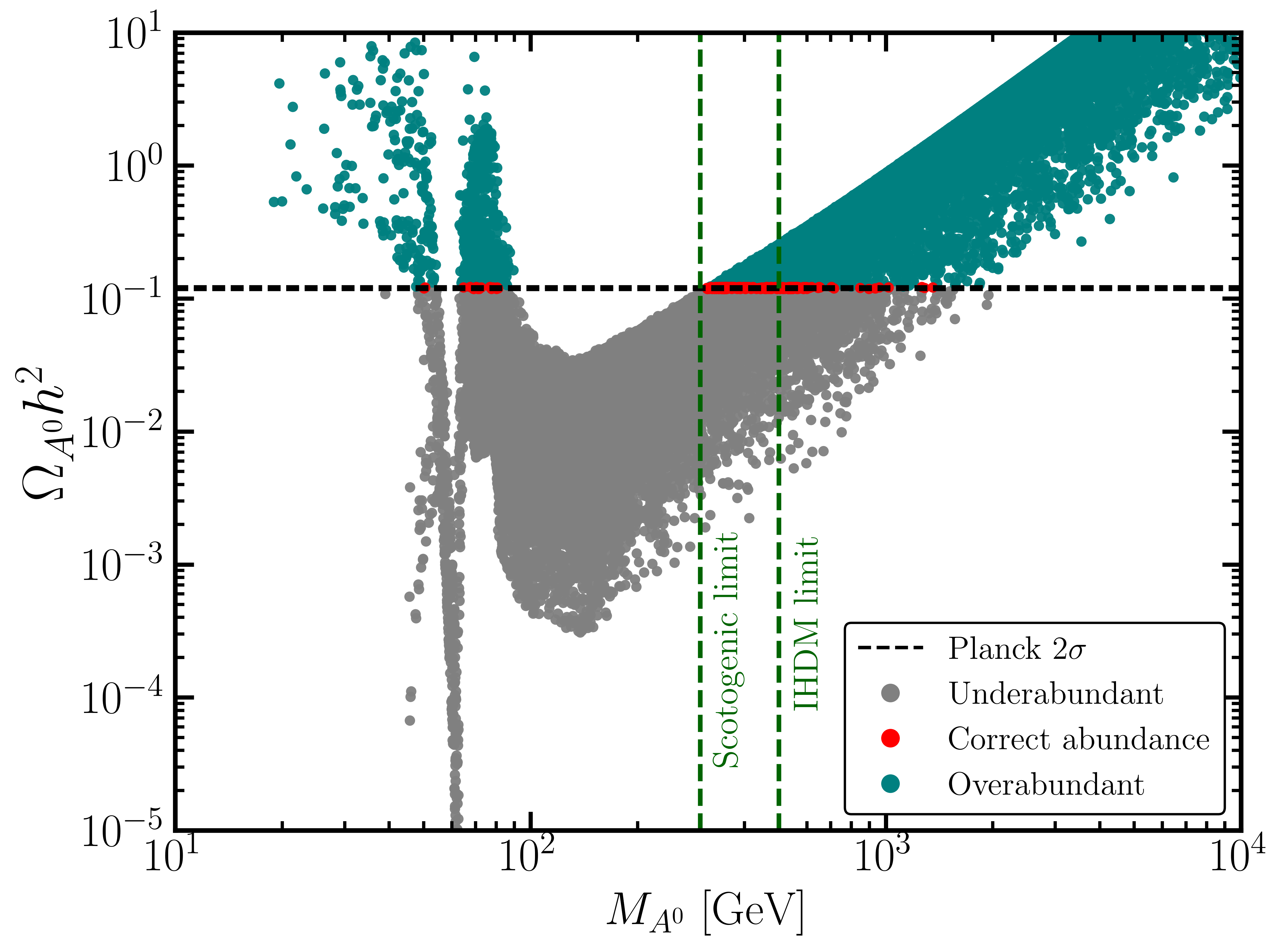}
    \caption{Parameter space of the DM relic abundance $\Omega_{A^0} h^2$ as a function of the DM mass $M_{A^0}$, including annihilation and co-annihilation channels for a radiation-dominated scenario, is shown.  The black dashed lines indicate the $2\sigma$ Planck relic density range. Gray, red, and cyan points correspond to underabundant, correct, and overabundant relic density, respectively. The vertical dashed green lines mark the IHDM and scotogenic limits.}
    \label{fig:relic_rad}
\end{figure}

The resulting prediction for the viable parameter space is shown in Fig.~\ref{fig:relic_rad}.
The DM relic density $\Omega_{A^0} h^2$, as a function of the DM mass $M_{A^0}$, is shown. The points yielding underabundant and overabundant relic densities are indicated in gray and cyan, respectively, while the points consistent with the observed relic density are highlighted in red. The horizontal dashed black lines denote the $2\sigma$ allowed range of $\Omega_{A^0} h^2$ \cite{Planck:2018vyg}. 
Furthermore, to emphasize the distinction between the relic density parameter spaces of the Scotogenic model and the IHDM, we have indicated two vertical green lines in the figure. 
The green line near $M_{A^0}\simeq 500$ GeV marks the lower mass threshold above which the observed DM relic abundance can be satisfied in the IHDM. It is important to emphasize that the dilution of DM either through annihilation into pairs of SM gauge bosons or (co-) annihilation into pairs of SM fermions through the SM (gauge) Higgs bosons remains highly efficient in the region $100 \leq M_{A^0} \leq 500$ GeV, resulting in DM underabundance. In contrast, the mass interval $100 \lesssim M_{A^0} \lesssim 500$ GeV, which is underabundant in the IHDM, can be brought into agreement with the observed relic density in the scotogenic model by considering all three $N_i$ states as co-annihilating partners \cite{Klasen:2013jpa}. However, increasing the number of co-annihilating fermions also enhances the effective annihilation rate relevant for indirect detection. Thus, although the inclusion of multiple co-annihilating partners can extend the viable relic density region towards lower DM masses, it is accompanied by an increased indirect detection cross-section and does not lead to a substantially larger viable parameter space once the current indirect detection constraints are imposed.
In our analysis, we therefore consider the more minimal scenario with only one $N_1$ acting as a co-annihilating partner. In this case, the observed relic density can be achieved down to $M_{A^0}\sim 300$ GeV while keeping the indirect detection cross-section well within the current experimental limits. The region below $M_{A^0}\sim300$ GeV, however, remains underabundant within the standard radiation-dominated cosmology. We find the value of $Y^\nu_{1\rm{eff}}$ in the range of $\mathcal{O}(10^{-5}-10^{-2})$ while being consistent with the neutrino oscillation data and condition provided in Eq.~\ref{eq:decaywidth_validity}.

An important point is that these conclusions are obtained under the assumption of the standard radiation-dominated thermal history of the Universe. As we demonstrate in the following subsection, this conclusion is substantially modified in a non-standard kination-dominated cosmological scenario, where the enhanced expansion rate during freeze-out significantly alters the predicted relic abundance and opens up additional viable regions of parameter space.

\FloatBarrier

\subsection{Kination Dominated Universe}

We now consider a non-standard cosmological history in which the Universe undergoes a kination-dominated phase prior to the onset of standard radiation domination. 
The additional kination component enhances the Hubble expansion rate, facilitating the early chemical decoupling of DM from the thermal bath. Consequently, the modified expansion history can provide the required enhancement in the DM relic density for underabundant DM parameter points and extend the viable mass range beyond that obtained in the standard radiation-dominated scenario.\footnote{Note that a DM overabundance predicted in the standard cosmological scenario will also generally result in an overabundance in the kination-dominated scenario. In such cases, the observed relic density can instead be realized through a matter-dominated epoch followed by entropy injection, which dilutes the DM abundance~\cite{Roy:2025moo}.}
In the presence of an additional component $\phi$, the total energy density of the Universe receives a contribution beyond the standard radiation bath, and the Hubble expansion rate is correspondingly modified. It can be expressed as
\begin{align}
H^2=\frac{\rho_R+\rho_\phi}{3M_P^2},
\end{align}
where $\rho_R$ and $\rho_\phi$ denote the radiation and kination energy densities, respectively.
The evolution of the $\phi$ component can be parametrized in terms of its equation-of-state parameter $w > 1/3$ as
\begin{align}
    \rho_\phi(T)=\rho_R(T_{RH})
\left(\frac{T}{T_{RH}}\right)^{3(1+w)},
\end{align}
where $T_{RH}$ denotes the reheating temperature, defined as the temperature at which the energy densities of the $\phi$ component and the radiation bath become equal,
$\rho_\phi(T_{RH})=\rho_R(T_{RH})$.
For $T>T_{RH}$, the $\phi$ component dominates over radiation and leads to an enhanced expansion rate compared to the standard radiation-dominated cosmology. As the Universe cools below $T_{RH}$, the radiation component becomes dominant, thereby recovering the standard cosmological evolution. The corresponding Hubble expansion rate can therefore be expressed as
\begin{align}
H(T)\simeq
\sqrt{\frac{\pi^2 g_{\rm eff}(T)}{90}}\,
\frac{T^2}{M_P} \times
\begin{cases}
\left(\dfrac{T}{T_{RH}}\right)^{\frac{3w-1}{2}},
& T\geq T_{RH},\\[6pt]
1,
& T\leq T_{RH}.
\end{cases}
\end{align}
The choice $w=1$ corresponds to the kination scenario, for which the energy density of the additional component scales as $\rho_\phi(T) \propto T^{6}$ \cite{Zeldovich:1961sbr,Spokoiny:1993kt,Joyce:1996cp,Ferreira:1997hj}. 

\begin{figure} [!t]
    \centering
    \includegraphics[width=0.48\linewidth]{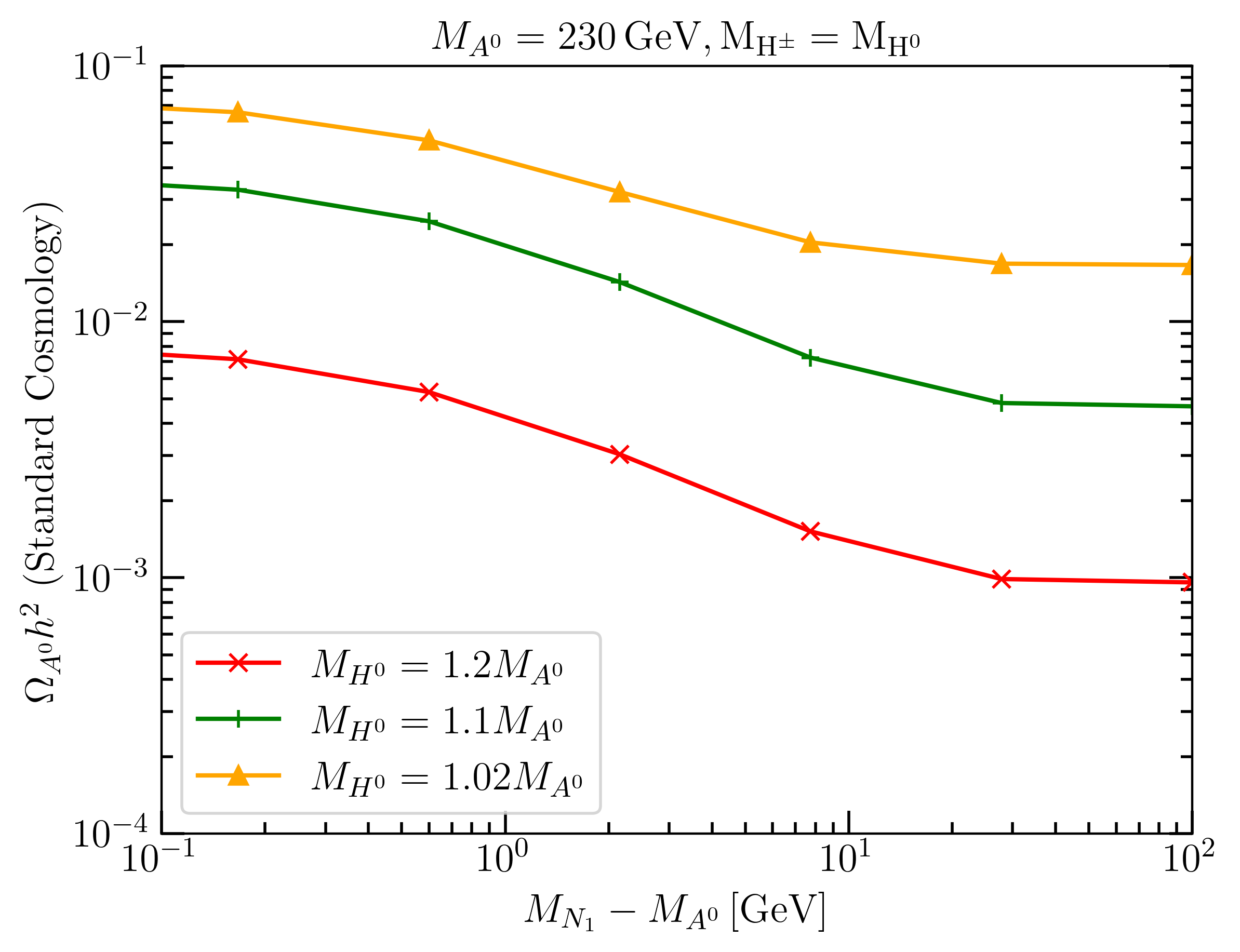}
    \includegraphics[width=0.48\linewidth]{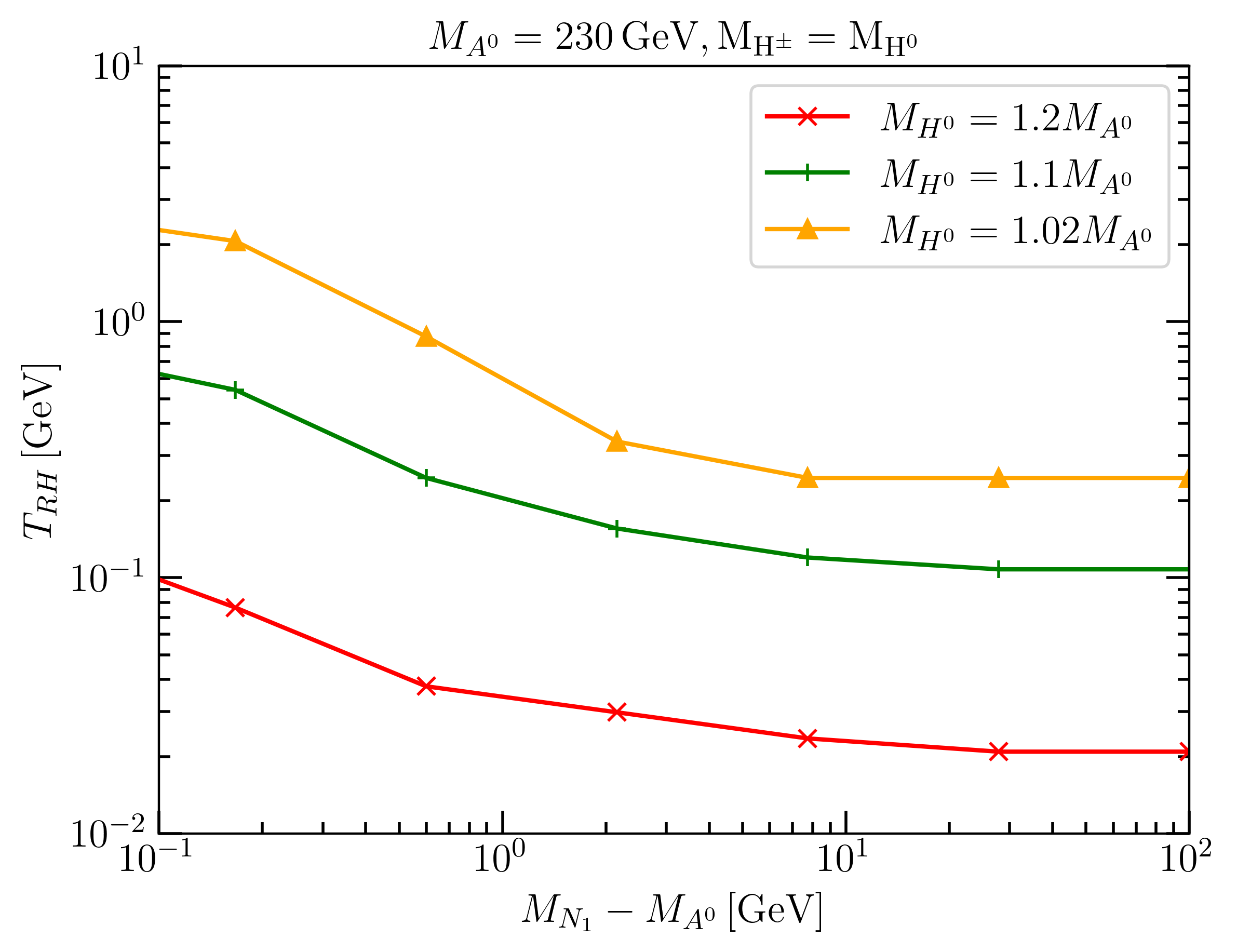}
    \includegraphics[width=0.48\linewidth]{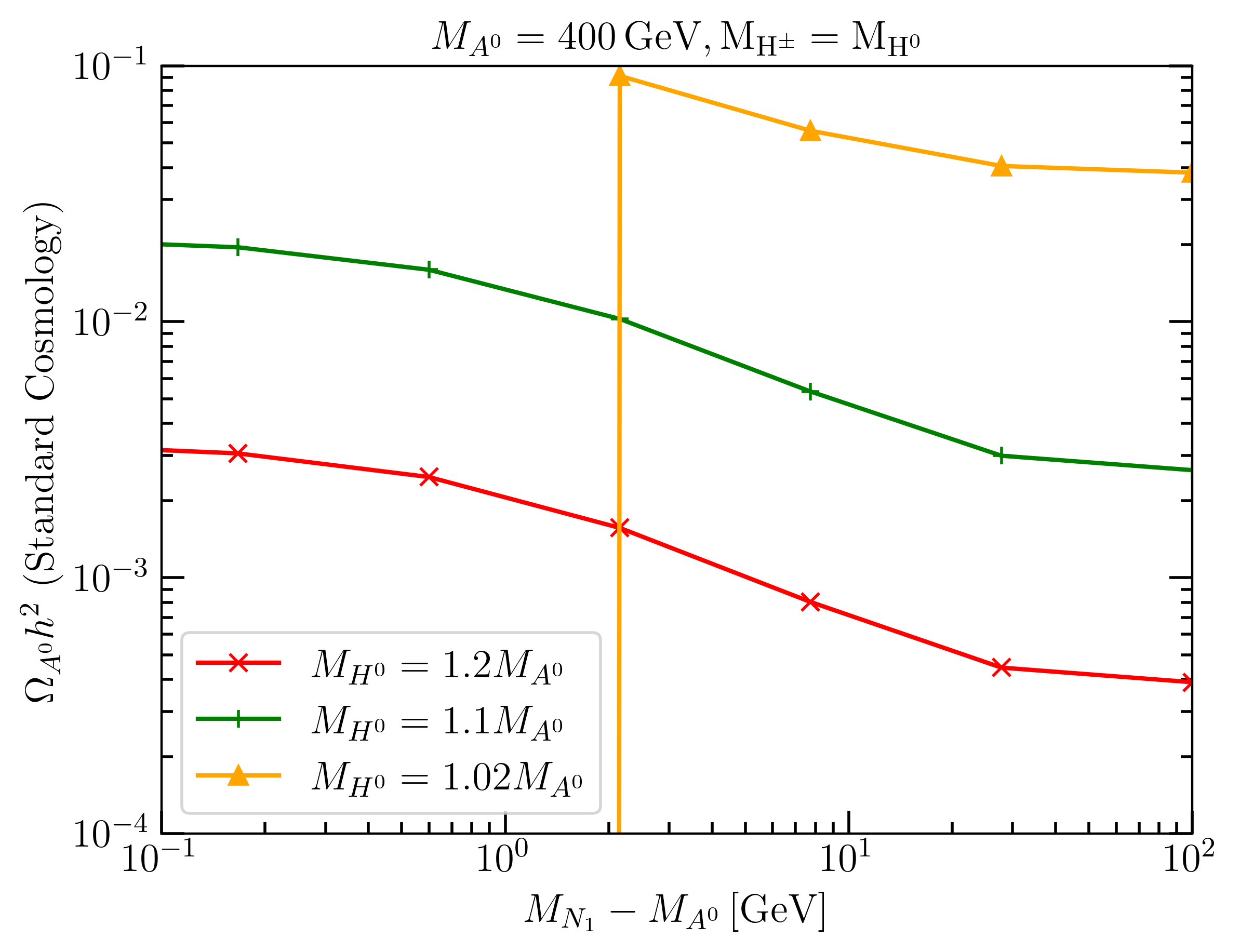}
    \includegraphics[width=0.48\linewidth]{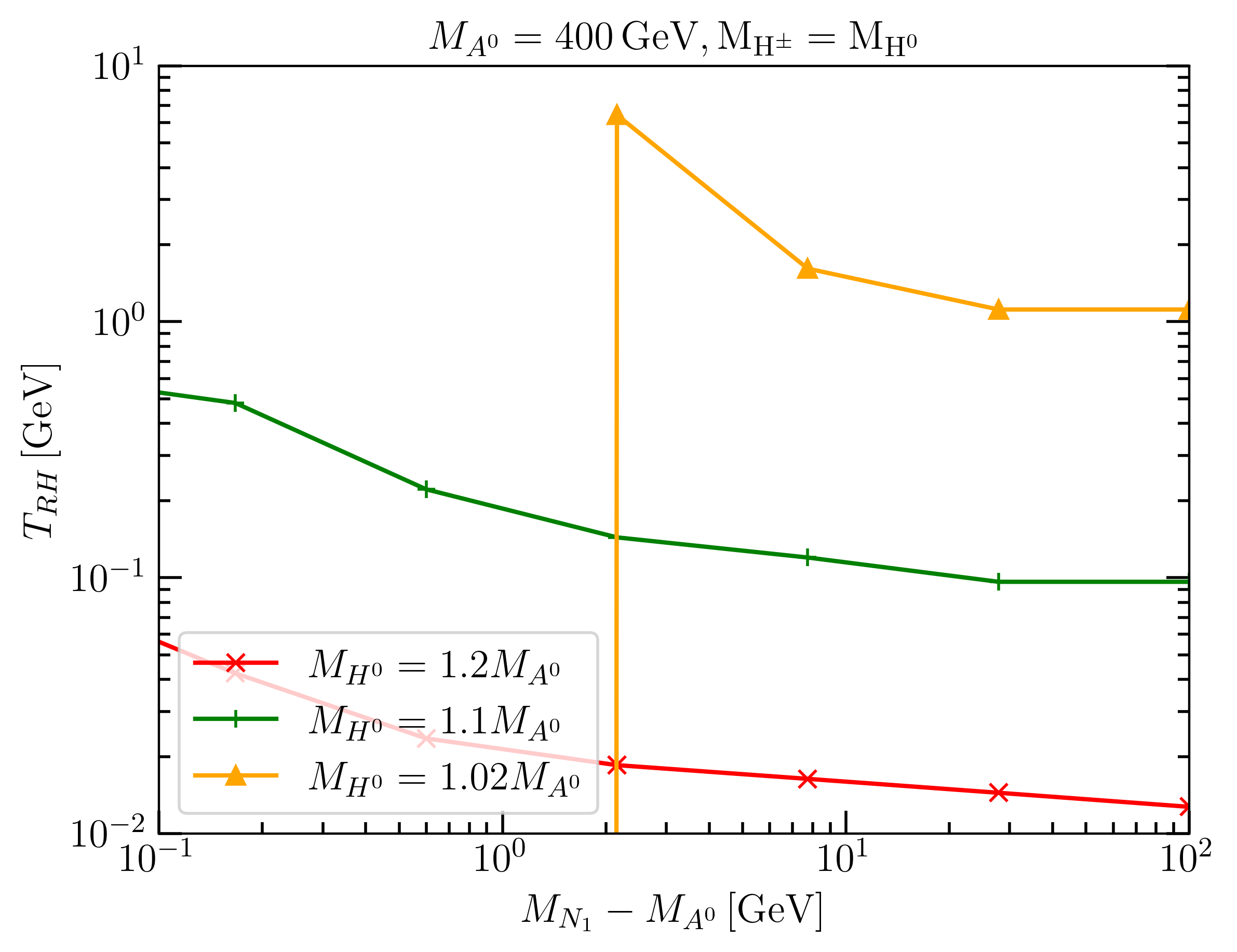}
    \caption{Correlation of the mass splitting \(M_{N_1}-M_{A^0}\) with the relic density in standard cosmology (left panels) and the reheating temperature \(T_{RH}\) (right panels), for \(M_{A^0}=230~\mathrm{GeV}\) (top panels) and \(400~\mathrm{GeV}\) (bottom panels). The orange, green, and red curves correspond to \(M_{H^0}=1.02M_{A^0}\), \(1.1M_{A^0}\), and \(1.2M_{A^0}\), respectively, with \(M_{H^0}=M_{H^\pm}\). In the right panels, the shown points satisfy the observed relic density constraint within \(2\sigma\) after accounting for the kination era. Note that we have also kept $M_{N_2}=5\,\rm{TeV}$, $M_{N_3}=10\,\rm{TeV}$ and $m_{\nu_1}=0.01\,\rm{eV}$ fixed.}
    \label{fig:kination}
\end{figure}
For the kination scenario, the $\phi$ energy density decreases more rapidly with the expansion of the Universe than the radiation energy density. Consequently, radiation later becomes the dominant component, ensuring the standard cosmological evolution required for BBN.  
During the kination period, the enhanced Hubble expansion rate drives the DM particles out of chemical equilibrium earlier than in the standard radiation-dominated cosmology. The earlier decoupling suppresses the subsequent annihilation efficiency and thereby leads to an enhancement of the final DM relic abundance.
This effect is particularly relevant for the underabundant parameter space, for which the kination-induced enhancement can bring the relic abundance into agreement with the observed DM density.
In our analysis, we therefore consider only the underabundant points in the DM mass range $100~{\rm GeV}\lesssim M_{A^0}\lesssim 1~{\rm TeV}$, identified within the standard radiation-dominated cosmology. For each of these parameter points, we determine the corresponding reheating temperature $T_{RH}$ required for the kination-induced modification of the cosmological evolution to reproduce the observed DM relic abundance in the $2\sigma$ limit. It is also important to draw the reader's attention to the fact that, for DM underabundant parameter points in the standard cosmological scenario corresponds to $\langle\sigma v\rangle_{eff} \gtrsim 3\times10^{-26}\,\mathrm{cm^3\,s^{-1}}$. Consequently, these parameter points are subject to stringent constraints from various direct and indirect DM detection searches and are likely to be further probed by upcoming experiments. We therefore impose the relevant theoretical, direct, and indirect detection constraints and identify the surviving parameter space by determining the appropriate reheating temperature $T_{\rm RH}$, with particular emphasis on future detection searches.

\begin{figure} []
    \centering
    \includegraphics[width=0.48\linewidth]{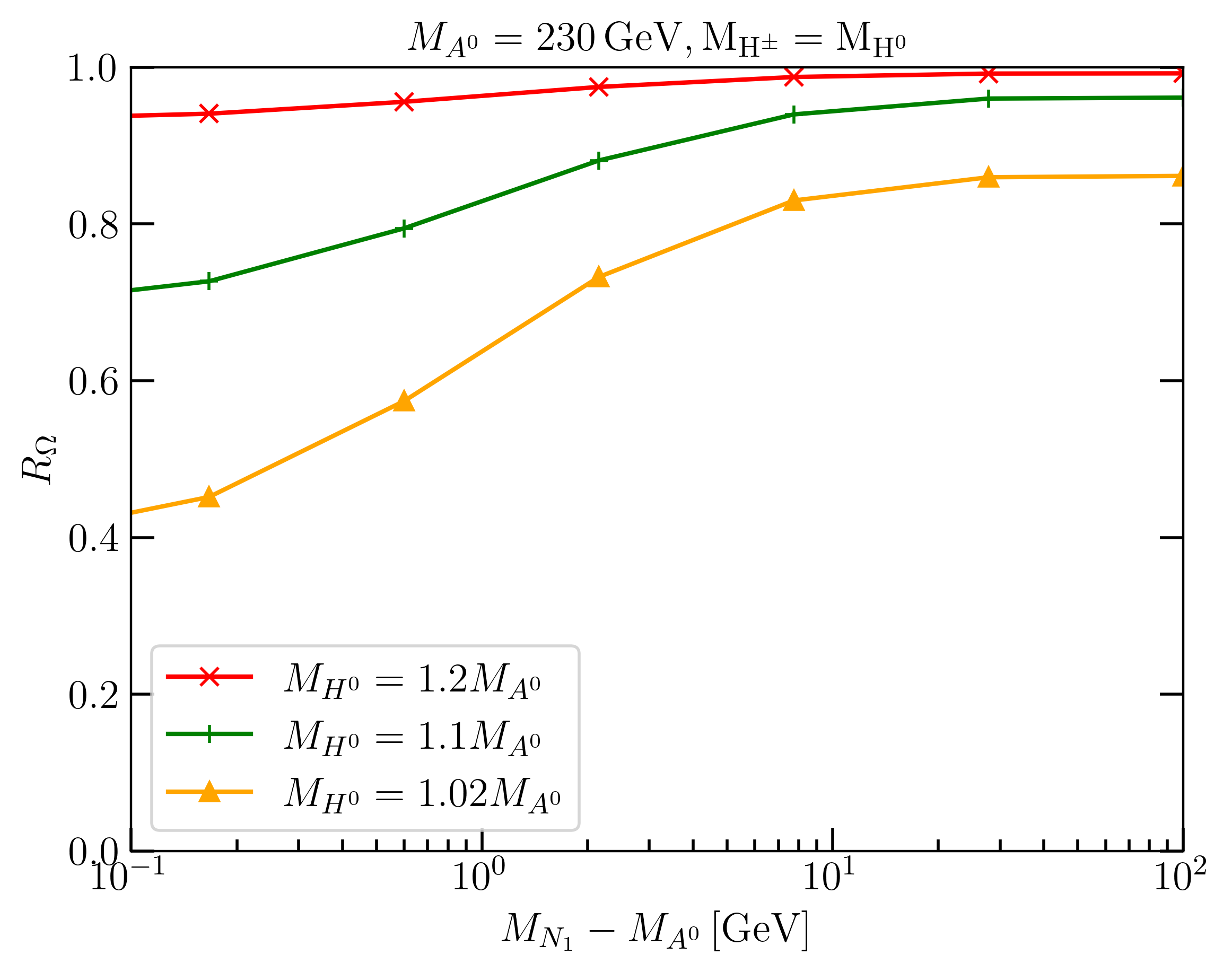}
    \includegraphics[width=0.48\linewidth]{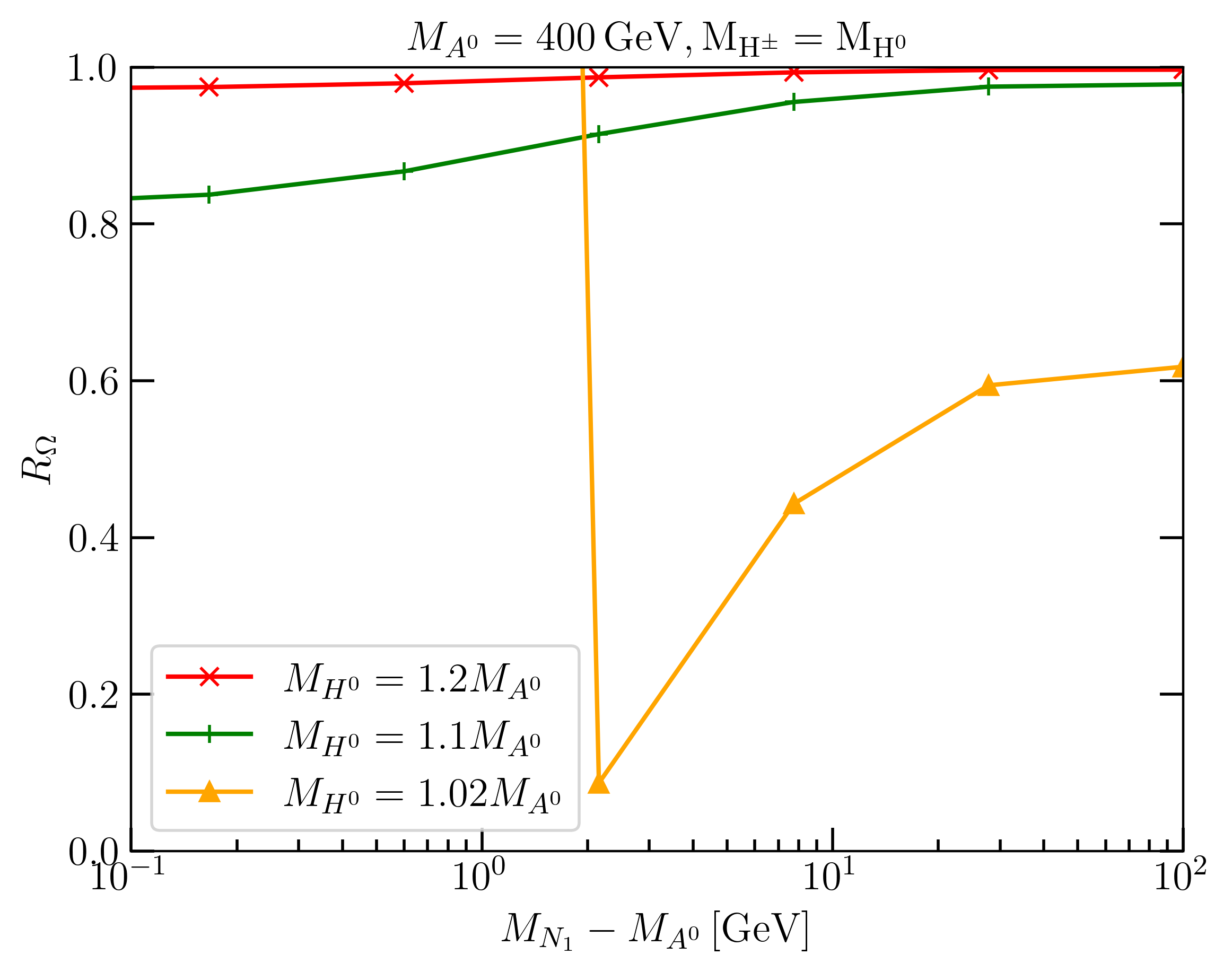}
    \caption{Interplay between the $R_{\Omega}$ and \(M_{N_1}-M_{A^0}\) for the same benchmark values of DM mass as in Fig.~\ref{fig:kination}. Note that all the points satisfy the observed relic density constraint within $2\sigma$ range after accounting for the
kination era.}
    \label{fig:romega}
\end{figure}

Before discussing the impact of kination on the various DM observables and the resulting viable parameter space, we first examine the role of scalar and fermionic co-annihilation, quantified by the mass splitting $(M_{N_1}-M_{A^0})$ and $(M_{H^\pm/H^0}-M_{A^0})$, respectively. The fermion splitting provides a useful distinction between the pure IHDM and the scotogenic scenario, where the additional fermionic state $N_1$ can participate in DM co-annihilation. We further explore its correlation with the reheating temperature $T_{RH}$, which characterizes the extent to which the kination era modifies the DM evolution. The results are shown in Fig. \ref{fig:kination}. The left~(right) panels show the correlation between the mass splitting and the relic density in standard cosmology~(kination-dominated scenario). We consider two benchmark values of the DM mass, $M_{A^0}=230~\mathrm{GeV}$ (top panel) and $M_{A^0}=400~\mathrm{GeV}$ (bottom panel). To show the effect of scalar co-annihilation, we take $M_{H^\pm}=M_{H^0}$ and consider following three choices: $M_{H^0}=1.02M_{A^0}$ (orange), $1.1M_{A^0}$ (green) and $1.2M_{A^0}$ (red).\footnote{These benchmark choices, together with $M_{H^\pm}=M_{H^0}$, are adopted solely for illustrative purposes to demonstrate the behaviour of the relic density as a function of $M_{N_1}-M_{A^0}$ and $T_{RH}$. They are not imposed in the full parameter scan, where these masses are allowed to vary independently, subject to the model constraints.} 
These benchmark choices correspond to different mass splittings among the inert scalars. 
The red curve in each panel stands for the no scalar co-annihilation scenario and moving from the red to the orange curve shows the effect of scalar co-annihilation. The increase in the mass splitting corresponds to a larger value of $\bar{\lambda}_L$, which controls the Higgs-portal interaction. Thus, moving from the red to the orange curve decreases the strength of the Higgs-portal interaction, leading to an increase in the relic density. In each panel, the dependence on $M_{N_1}-M_{A^0}$ further captures the contribution from fermionic co-annihilation with $N_1$. 

\par The upper left panel of Fig.~\ref{fig:kination} shows the correlation of relic density with mass splittings in the standard cosmology for dark matter mass $M_{A^0}=230$~GeV, which reconfirms the results of Fig.~\ref{fig:relic_rad}.  The red line shows the effect of only fermion co-annihilation, whereas the green and orange lines show the effect of both scalar and fermion co-annihilation. It shows that both fermion and scalar co-annihilation help to increase the relic density but still not enough to reproduce the correct relic density for this DM mass. As the mass splitting increases, these processes become progressively Boltzmann suppressed, and the relic density decreases accordingly. For sufficiently large \(M_{N_1}-M_{A^0}\), the contribution of \(N_1\) co-annihilation becomes negligible, and the result approaches that of the pure IHDM.

\par For the same benchmark choices, the right panels illustrate the correlation between scalar/fermion mass splitting and $T_{\rm RH}$, required to bring the underabundant points into the \(2\sigma\) range of the observed relic density. We find that the smaller the scalar/fermion mass splitting, the larger the required \(T_{RH}\), indicating that only a small modification of the standard cosmological evolution is required. As the scalar or fermion mass splitting increases, co-annihilation becomes progressively less effective, the required \(T_{RH}\) decreases. This reflects a clear interplay between co-annihilation and the kination era: when co-annihilation is efficient, the relic density is already strongly affected by the additional annihilation channels, and the impact of kination can therefore remain small. In contrast, when co-annihilation is suppressed, a lower \(T_{RH}\) and hence a stronger kination effect is required to bring the relic density into agreement with observations.


%
\begin{figure} []
    \centering
\includegraphics[width=0.55\linewidth]{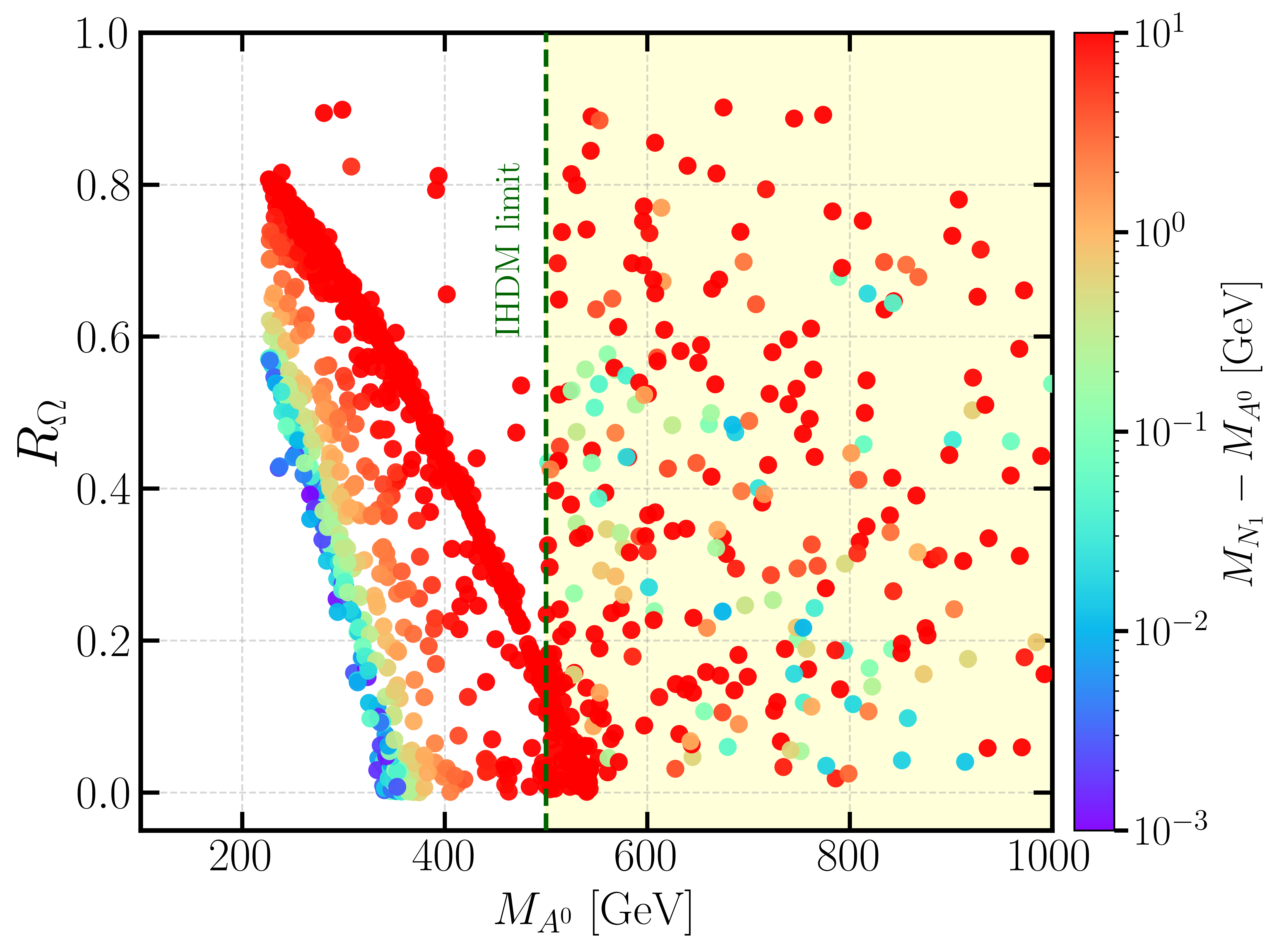}
\caption{Impact of the kination cosmology on the viable DM parameter space. $R_{\Omega}$ as a function of $M_{A^0}$ is shown, color-coded according to the mass splitting $(M_{N_1}-M_{A^0})$.
The vertical dashed green line corresponds to the IHDM limit and the yellow-shaded region indicates the parameter space where the observed relic density can be obtained without requiring co-annihilation with the dark fermion $N_1$.
Note that all points agree with the best-fit neutrino oscillation data and the observed DM relic density within the $2\sigma$ limit.}
    \label{fig:coanni}
\end{figure}

Furthermore, to quantify the relevance of this (co-)annihilation mechanism and the impact of the kination-dominated cosmology, we introduce the following quantity: 
\begin{align}
    R_{\Omega}= 1- \frac{\Omega_{A^0}h^2(T_{RH}\gg T_{dec})}{\Omega_{\rm DM}^{\rm obs}h^2},
\end{align}
Here, $\Omega_{A^0}h^2(T_{RH}\gg T_{dec})$ denotes the relic abundance of the DM candidate $A^0$ calculated within the standard radiation-dominated cosmology for the parameter points subsequently considered in the kination regime with $T_{RH}<T_{dec}$, while $\Omega_{\rm DM}^{\rm obs}h^2$ represents the observed DM relic density.
For $T_{RH}\gg T_{dec}$, the DM particle decouples after the transition to the radiation-dominated era, and the standard thermal freeze-out scenario is recovered. For the underabundant parameter points considered here, this corresponds to $R_{\Omega}\to1$ when the standard cosmological relic abundance is much smaller than the observed DM density. Conversely, $R_{\Omega}\to0$ corresponds to parameter points whose relic abundance in the standard radiation-dominated cosmology is already close to the observed DM density, and hence requires only a small kination-induced enhancement to reproduce the observed relic abundance.
To illustrate the interplay between \(R_{\Omega}\) and the mass splitting \(M_{N_1}-M_{A^0}\), we present the corresponding results in Fig. \ref{fig:romega} for the same DM mass benchmarks considered in Fig. \ref{fig:kination}. From these plots, we can infer that for the smaller mass splitting \(M_{N_1}-M_{A^0}\), we can have smaller values of $R_{\Omega}$ depending on the dark scalar's mass splitting. In the right panel, we can see that for $M_{H^0}=1.02 M_{A^0}$, $R_{\Omega}$ is close to zero, showing the small effect of kination. However, from both panels, we can observe that as \(M_{N_1}-M_{A^0}\) increases, the co-annihilation contribution becomes progressively suppressed, and \(R_{\Omega}\) approaches 1, indicating that a substantial modification of the relic density through the kination era is required.

To examine the impact of the kination era across different DM mass regimes, we present the full parameter space analysis in Fig. \ref{fig:coanni}. The figure shows the variation of \(R_{\Omega}\) with the DM mass \(M_{A^0}\), with the points color-coded according to the mass splitting \(M_{N_1}-M_{A^0}\). We would like to mention that all the shown points satisfy the correct relic density in the $2\sigma$ range and are consistent with indirect and direct detection constraints, as we discuss in the next Sec. \ref{sec:dmobs}.
We find that regions with larger $R_{\Omega}$, corresponding to points requiring a stronger kination-induced enhancement of the relic abundance, are mainly concentrated towards lower values of DM mass $M_{A^0}$. 
From the figure, it is also evident that below $300$ GeV of DM mass, we always need kination, i.e., $R_{\Omega}>0$. This region also corresponds to the lower values of the mass difference $M_{N_1}-M_{A^0}$ and hence the stronger co-annihilation effect. 

To further illustrate this behavior, we shade in yellow the region up to the IHDM limit, corresponding to the range of DM masses where the observed relic density can be obtained without requiring co-annihilation from the dark fermion. In the IHDM, such a contribution from the dark fermion is absent, providing a useful reference for our scenario. Within this shaded region, the viable points exhibit no pronounced correlation between $M_{N_1}-M_{A^0}$ and the kination effect, indicating that the relic abundance can be obtained without a significant contribution from $N_1$ co-annihilation. Below the IHDM limit, however, a mild correlation becomes apparent, suggesting that both the kination-induced modification of the relic abundance and co-annihilation with $N_1$ play a role in determining the viable parameter space.

\section{Dark Matter Observables} \label{sec:dmobs}

\subsection{Indirect Detection}

We now discuss the implications of indirect detection searches for the DM candidate $A^0$.
The production rate of particle $\kappa$ (here $\gamma,e^{-}\ \text{or}\ \bar{p}$) from $A^{0}$ annihilation is given by~\cite{Belanger:2021lwd,Belanger:2022qxt,Belanger:2024wca},
\begin{eqnarray}
Q^{\kappa}(E)=\frac{1}{2}\frac{\rho_{A^{0}}^{2}}{M_{A^{0}}^2}\langle\sigma v\rangle\frac{dN^{\kappa}(E)}{dE},
\label{eq:ID_DM2}
\end{eqnarray}
where $\frac{dN^{\kappa}(E)}{dE}$ is the number of $\kappa$ produced per unit energy in  $A^{0}$ pair annihilation.

Note that in our analysis, we have considered combined $\kappa$ spectra from all DM annihilation channels to compute $Q^{\kappa}(E)$. Therefore, $\frac{dN^{\kappa}(E)}{dE}$ can be written as,
\begin{eqnarray}
\frac{dN^{\kappa}(E)}{dE}=\sum_{i}\text{BR}(A^{0}A^{0}\to ii)\left(\frac{dN^{\kappa}(E)}{dE}\right)_{i},
\label{eq:ID_DM3}
\end{eqnarray}
where $\left(\frac{dN^{\kappa}(E)}{dE}\right)_{i}$ is $\kappa$ spectra associated with annihilation of $A^{0}A^{0}$ into $ii$ final states and $\text{BR}(A^{0}A^{0}\to ii)$ corresponds to fraction of time $A^{0}$ annihilation will end up into final state $i$. The branching fraction $\text{BR}(A^{0}A^{0}\to ii)$, therefore, can be written as,
\begin{eqnarray}
\text{BR}(A^{0}A^{0}\to ii)=\frac{\langle \sigma v \rangle_{A^{0}A^{0}\to ii}}{\langle\sigma v\rangle},
\label{eq:ID_DM4}
\end{eqnarray} 
where $\langle\sigma v\rangle$ is the thermally averaged annihilation  cross-section of $A^0$ pairs,  taking into account all the annihilation channels.
It is also important to highlight that no co-annihilating partner of $A^{0}$ contributes to $\langle\sigma v\rangle$ as it has decayed to SM particles at the early epoch of the Universe.

We then use \texttt{micrOMEGAs7.1} to determine the indirect detection constraints derived from observations of dwarf spheroidal galaxies by the Fermi-LAT telescope \cite{Alvarez:2020cmw}. Furthermore, the photon spectra are computed using \texttt{micrOMEGAs7.1}, and the analysis presented in Ref.~\cite{CTA:2020qlo} is followed to estimate the expected sensitivity of the CTA for our model parameter space.

\begin{figure} []
    \centering
    \includegraphics[width=0.55\linewidth]{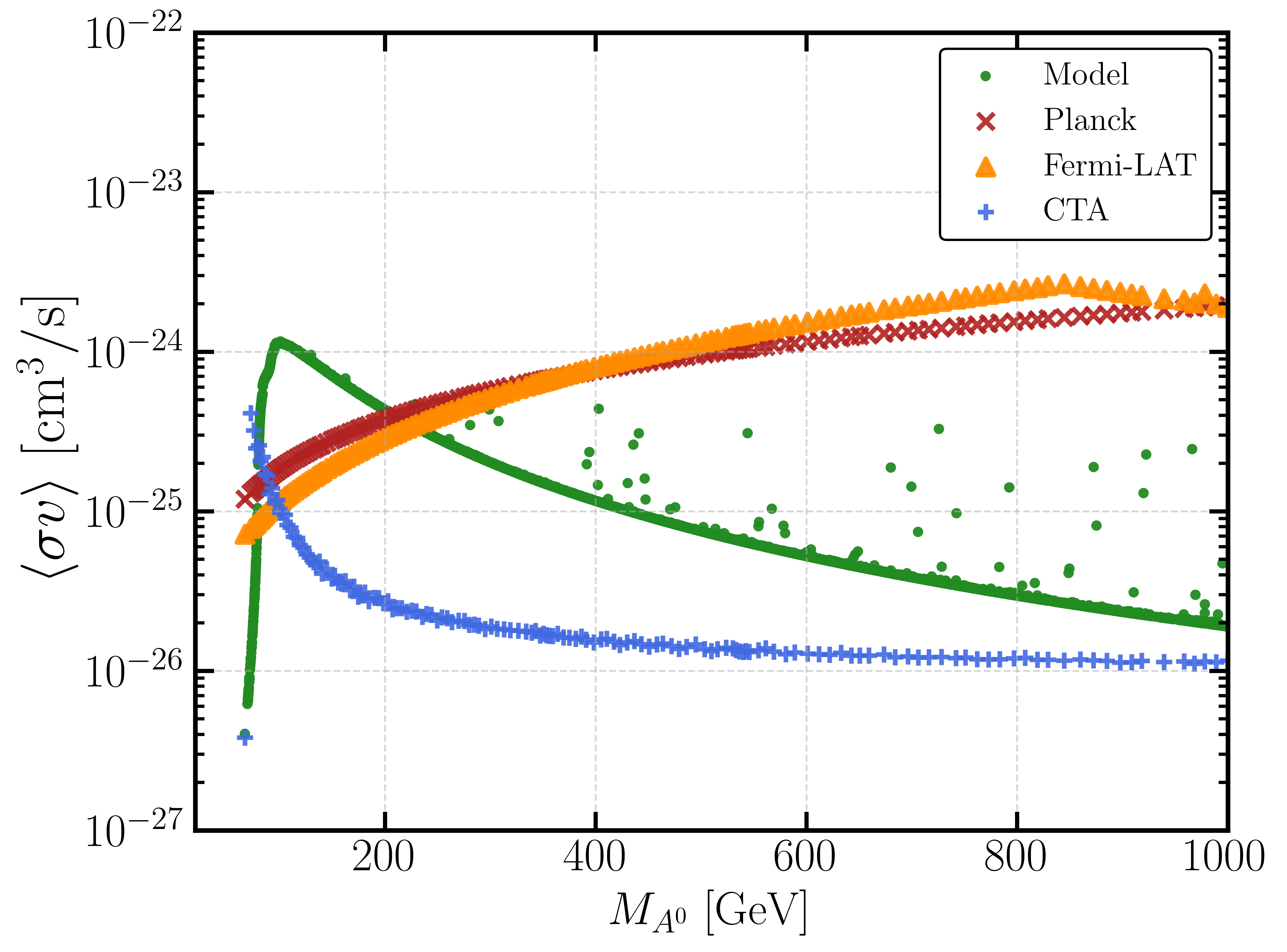}
    \caption{Indirect detection constraints and future prospects for the DM candidate $A^0$. The green points show the predicted DM annihilation cross-sections for parameter points. The current indirect detection limits from Planck (red) and Fermi-LAT (orange), along with the projected sensitivity of CTA (blue), are shown for comparison. Note that all the points satisfy the observed DM relic abundance within the $2\sigma$ limit.}
    \label{fig:indirect1}
\end{figure}

\subsection{Cosmic Microwave Background}
The Cosmic Microwave Background (CMB) is sensitive to DM annihilation around the period of the recombination epoch. The energy injection from DM annihilation to SM particles into the photon-baryon plasma during the recombination epoch can affect the ionization history and the temperature~\cite{Lopez-Honorez:2013cua,Diamanti:2013bia,Batell:2017rol,Finkbeiner:2011dx,Slatyer:2015jla}. The energy injection per unit volume per unit time at redshift $z$ is given by,
\begin{equation}
\frac{dE}{dV dt} = \rho_{\rm c}^2\, \Omega_{\chi}^2\, (1+z)^6 \left[f(z) \frac{\langle \sigma v \rangle}{M_{A^{0}}} \right],
\label{eq:CMB}
\end{equation}
where $\Omega_{\chi}$ is the observed DM density, $\rho_{c}$ corresponds to the critical density of the Universe, and $\langle \sigma v \rangle$ is the DM thermal average cross-section around the period of the recombination epoch. The efficiency factor, denoted as $f(z)$, accounts for the fraction of injected energy deposited into the intergalactic medium (IGM) at redshift $z$. This factor depends on the spectrum of photons and electrons from the DM annihilations. The efficiency factor $f(z)$ can be approximated by a constant parameter referred to as $f_{\text{eff}}$. The validity of this approximation is due to the CMB data being susceptible to energy injection within a narrow redshift range, i.e., ($1000$-$600$)~\cite{Slatyer:2009yq}. The $f_{\text{eff}}$ as function of DM mass, $M_{A^{0}}$, is given by,
\begin{eqnarray}
f_{\text{eff}}(M_{A^{0}})=\frac{1}{2 {M_{A^0}}} \displaystyle{\int_{0}^{M_{A^{0}}}\!\! dE\, E \left[2f_{\text{eff}}^{e^-}(E)\left(\frac{dN}{dE}\right)^{e^-} + f_{\text{eff}}^{\gamma}(E)\left(\frac{dN}{dE}\right)^\gamma\right]},
\label{eq:feff}
\end{eqnarray}
where $(dN/dE)^{\gamma, e^-}$ corresponds to the photon/electron spectrum and $f^{\gamma (e^{-})}_{\text{eff}}$ coefficients as a function of energy results are obtained from Ref.~\cite{Slatyer:2015jla}. The Planck CMB measurement sets the following limit~\cite{Planck:2018vyg}
\begin{eqnarray}
p_{ann}=f_{\text{eff}}(M_{A^{0}})\frac{\langle\sigma v\rangle}{M_{A^{0}}} < 3.2 \times10^{-28}\, \text{cm}^3 \,\text{s}^{-1}\, \text{GeV}^{-1}.
\label{eq:planck}
\end{eqnarray} 
We have considered effective photon/electron spectra (see Eq.~\ref{eq:ID_DM3}) from DM annihilation into all SM states to compute $f_{eff}$, followed by determining the predicted $p_{ann}$ to find viable parameter space for our model.    

We present $\langle\sigma v\rangle$ as a function of DM mass $M_{A^0}$ in Fig. \ref{fig:indirect1}. We show the model predictions for the DM annihilation cross-section as green points and compare them with the current indirect detection constraints from Planck and Fermi-LAT, represented by the red and orange curves, respectively. The projected sensitivity of the CTA is shown by the blue curve. All the points satisfy the observed DM relic abundance within the $2\sigma$ limit.
The comparison provides a direct assessment of the present and future sensitivity of indirect detection experiments to the viable DM parameter space of the model.
As evident from the figure, the region with DM mass range $100 \leq M_{A^0} \leq 200$ is strongly constrained by the current indirect detection data, with the predicted annihilation cross-sections lying above the corresponding Planck and Fermi-LAT limits. This leads to the exclusion of a substantial fraction of the parameter space in this low-mass region.
Interestingly, the projected CTA sensitivity extends considerably into the parameter space that is currently allowed. In particular, a substantial fraction of the model predictions in the mass range $100 \leq M_{A^0} \leq 1$ TeV lies within the expected reach of CTA. This suggests that future CTA observations could significantly improve upon the present indirect detection constraints and probe regions of the DM parameter space that remain unconstrained by Planck and Fermi-LAT.

\subsection{Direct Detection}

We now examine the direct detection prospects of the DM candidate $A^0$. 
At tree-level, the spin-independent (SI) scattering of $A^0$ off nucleons receives contributions from two processes involving the quark constituents of the nucleon. One is mediated by the $Z$ boson, while the other proceeds through Higgs $h$ exchange. 
These interactions give rise to the corresponding spin-independent DM-nucleon scattering cross-sections. 
The SI cross-section associated with inelastic scattering between $A^0$ and nucleons through $Z$ boson exchange is given by
\begin{align}
\sigma_{Z}^{\rm SI}=\frac{G_{F}^{2}\mu_{r}^2}{2\pi}\left[N-(1-4\sin^2{\theta_W})Z\right]^2,
\end{align}
where $G_{F}$ is the Fermi constant, $\theta_{W}$ is the weak mixing angle, $\mu_{r}=\frac{M_{A^0}M_{N}}{M_{A^0}+M_{N}}$($M_{N}$ is the nucleon mass), and $N$ and $Z$ correspond to the number of neutrons and protons in the nucleus, respectively. 
The current experimental limits from direct detection experiments put severe constraints on $\sigma_{Z}^{\rm SI}$. Therefore, we assume the mass splitting between $M_{H^0}$ and $M_{A^0}$ induced by $\lambda_{5}$ is larger than the kinetic energy of DM in our galactic halo~\cite{Barbieri:2006dq}. 
Therefore, the scattering of $A^0$ on a nucleon through $Z$ boson exchange is kinematically forbidden. 
Consequently, the tree-level $h$ exchange between $A^0$ and a nucleon facilitated by the Higgs portal coupling $\overline{\lambda}_{L}$ gives the dominant contribution to the SI cross-section. The SI cross-section mediated through the Higgs portal is given by~\cite{Garcia-Cely:2015khw},
\begin{eqnarray}
\sigma^{\rm SI}_{h}=\frac{M_{N}^{4}f^{2}\overline{\lambda}_{L}^2}{4\pi(M_{N}+M_{A^0})^{2}M_{h}^4},
\label{Eqn:SI_DD}
\end{eqnarray}
where $f\approx0.3$ is a nucleon form factor. 

\begin{figure} [!t]
    \centering
    \includegraphics[width=0.55\linewidth]{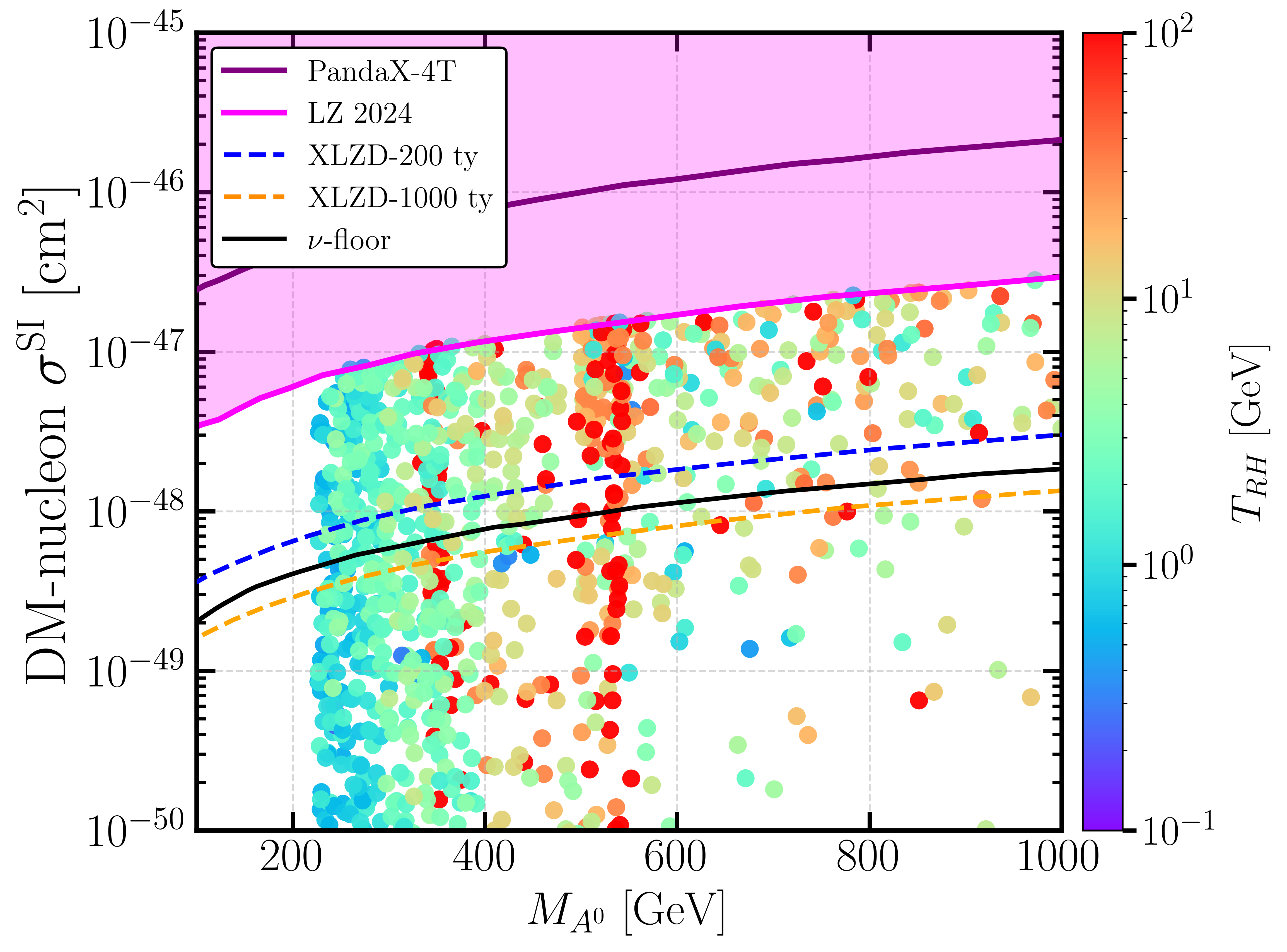}
    \caption{The spin-independent DM–nucleon cross-section, $\sigma^{\rm SI}$, as a function of the DM mass $M_{A^0}$. The scattered points are color-coded according to the reheating temperature $T_{\rm RH}$.
    The current exclusion limits from LZ (magenta) and PandaX-4T (purple), together with the projected XLZD sensitivities for 200 and 1000 tonne-years (dashed blue and orange, respectively), are shown. The black curve denotes the neutrino floor. Note that all points agree with the best-fit neutrino oscillation data and the observed DM relic density within the $2\sigma$ limit.}
    \label{fig:DD}
\end{figure}

We present our result for direct detection of the DM candidate $A^0$ in Fig. \ref{fig:DD}.
In this plot, the SI DM-nucleon cross-section $\sigma^{\rm SI} \approx \sigma^{\rm SI}_h$ as a function of the DM mass $M_{A^0}$ has been shown. The color coding of the scattered points represents the corresponding reheating temperature, $T_{RH}$. 
The current exclusion limits from PandaX-4T \cite{PandaX:2024qfu} and LZ \cite{LZ:2024zvo} are drawn by purple and magenta solid curves, respectively. The future XLZD \cite{XLZD:2024nsu} sensitivities for 200 and 1000 tonne-years are shown by blue and orange dashed curves, respectively. In addition, we have shown the neutrino floor \cite{Billard:2013qya} with a solid black line. 
The parameter space with $M_{A^0}\lesssim 200$ GeV is already excluded by the current indirect detection constraints.
The scattered points correspond to parameter points that reproduce the observed relic density within the $2\sigma$ range after accounting for the modified cosmological evolution due to kination, while remaining below the LZ exclusion limit. These points are therefore consistent with the current direct detection constraints. In the low $M_{A^0}$ regime, particularly for $M_{A^0}\lesssim 300$ GeV, the viable points are predominantly associated with relatively low values of $T_{RH}$. This indicates that lower values of $T_{RH}$, corresponding to a stronger kination effect, are required to obtain the observed DM relic abundance in this region of the parameter space.
In contrast, in the high $M_{A^0}$ regime, viable points with the correct relic density are already obtained within the standard cosmological scenario (see Fig. \ref{fig:relic_rad}). Moreover, a substantial fraction of the underabundant points lie close to the observed $2\sigma$ relic density range, and therefore require only a mild modification of the relic abundance from the kination-dominated cosmology. Consequently, this region of the parameter space can accommodate a broad range of reheating temperatures, from relatively low to high $T_{RH}$ values, as shown in Fig. \ref{fig:DD}.

\FloatBarrier


\section{Collider Observables}
\begin{figure}[]
\centering
\includegraphics[height=3.3cm, width=4.5cm]{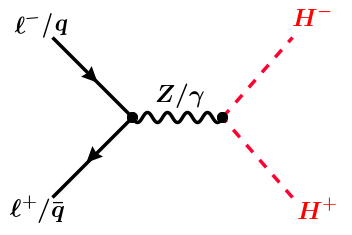}~~
\includegraphics[height=3cm, width=3.5cm]{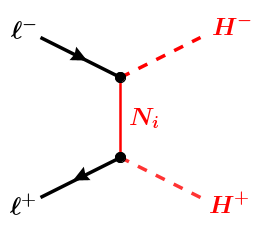}~~
\includegraphics[height=3.3cm, width=4.5cm]{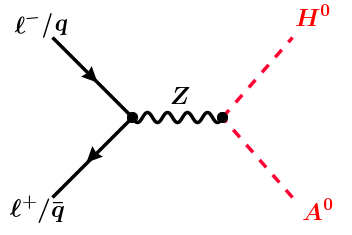}~~
\includegraphics[height=3.3cm, width=4.5cm]{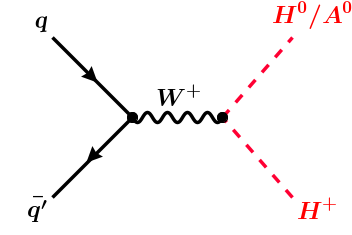}
\caption{
Dark scalar production channels at $pp$ and lepton colliders which can be produced by neutral- or charged-current Drell-Yan mechanisms. 
At lepton colliders, charged scalars can also be produced by t-channel dark fermion exchange.} 
\label{fig:production-scotogenic-1}
\end{figure}

\begin{figure} []
\centering
\includegraphics[width=0.48\linewidth]{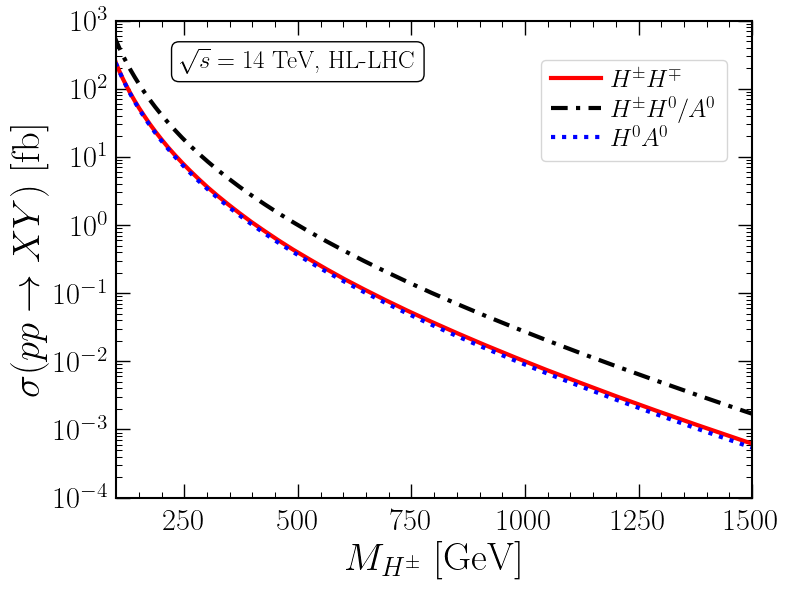}
\includegraphics[width=0.48\linewidth]{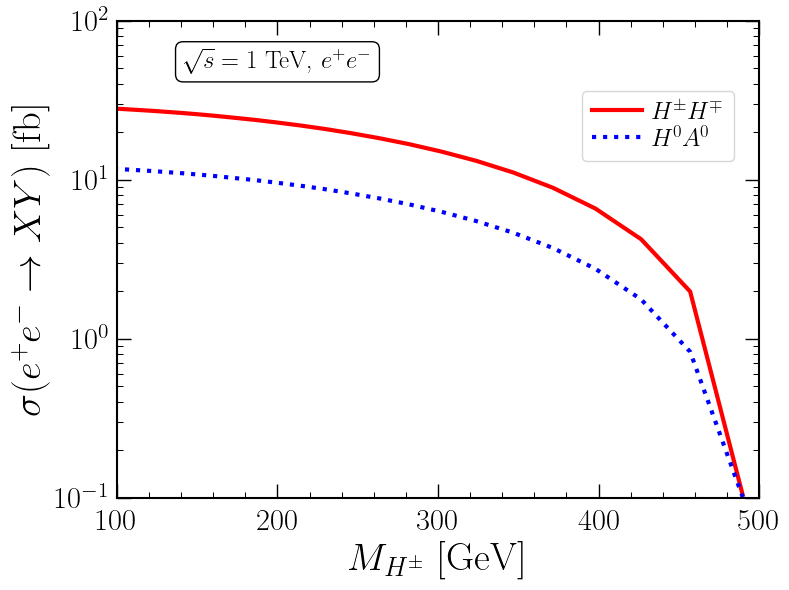}
\caption{Production cross-section of dark scalars at hadron~(left panel) and lepton collider~(right panel) as a function of charged scalar mass.}
\label{fig:XS}
\end{figure}

The $Z_2$-odd particles introduced in the IHDM and scotogenic model can give rise to observable signatures at hadron colliders, such as the LHC and the proposed FCC-hh, as well as at future lepton colliders, including CEPC~\cite{CEPCStudyGroup:2018ghi,Antusch:2025lpm}, ILC~\cite{Barklow:2015tja}, CLIC~\cite{CEPCStudyGroup:2018ghi,FCC:2018evy,Bambade:2019fyw,CLIC:2018fvx,Antusch:2025lpm}, and muon colliders~\cite{Delahaye:2019omf,Li:2023tbx,InternationalMuonCollider:2025sys,Stratakis:2025ofm}. In particular, the electroweak scale dark sector states associated with radiative neutrino mass generation may be directly produced at these facilities. At both hadron and lepton colliders, the inert scalars can be produced through neutral- and charged-current Drell-Yan processes, as illustrated in Fig.~\ref{fig:production-scotogenic-1}. In Fig.~\ref{fig:XS}, we show the corresponding production cross-sections as functions of the charged-scalar mass $M_{H^\pm}$ at the $\sqrt{s}=14$ TeV HL-LHC~\cite{Apollinari:2015wtw} and at a $\sqrt{s}=1$ TeV lepton collider. The production rates can remain sizable over a substantial part of the parameter space considered in this work.\footnote{At lepton colliders, $t$-channel exchange of the singlet fermions $N_i$ provides an additional contribution to $\ell^+\ell^-\to H^+H^-$. The pure $t$-channel contribution scales as $|Y^{\nu}_{i \alpha}|^4$ and can dominate over the Drell-Yan contribution for sufficiently large Yukawa couplings~\cite{Batra:2023mds,Batra:2023ssq,Chianese:2021toe}.}

\subsection{IHDM}
We first consider the collider phenomenology of the IHDM, which provides a useful reference for the scotogenic scenario. The gauge-mediated production mechanisms of the inert scalars are identical in the two models. The principal difference is the presence of the singlet Majorana fermions $N_i$ and their Yukawa interactions,
$Y^\nu\overline{N}\widetilde{\Phi}^{\dagger}L$, in the scotogenic model. In the scalar DM scenario considered throughout this work, the mass hierarchy is $M_{A^0}<M_{H^0},M_{H^\pm}$,
so that $A^0$ is the lightest $Z_2$-odd particle and constitutes the DM candidate. Since $A^0$ escapes detection, the generic collider signatures involve missing transverse energy accompanied by visible particles arising from initial-state radiation or from the decays of the heavier inert scalars. Representative signatures are $\slashed{E}_T+X$, where $X$ may consist of one or two jets~\cite{ATLAS:2016bek,Haisch:2018hbm}, two leptons~\cite{ATLAS:2018lkz}, or a photon~\cite{ATLAS:2017nga}. A particularly useful channel is the monojet signature, which can arise from
\begin{equation}
pp\to H^0A^0+j,\qquad
pp\to H^\pm(H^0/A^0)+j,\qquad
pp\to H^+H^-+j\, ,
\end{equation}
after the heavier inert states decay into the DM candidate~\cite{ATLAS:2016bek,ATLAS:2016neq}. In the radiation-dominated IHDM reference scenario relevant to our analysis, the observed relic abundance is predominantly obtained in the high-mass regime, $M_{A^0}\gtrsim500$ GeV. The corresponding electroweak production cross-sections are therefore typically smaller than those in the lower-mass regions opened by fermionic co-annihilation or kination.
\subsection{Scotogenic model: Radiation-dominated cosmology}
The scotogenic model offers additional collider possibilities owing to the presence of the singlet fermions $N_i$~\cite{Avila:2021mwg}. In the standard radiation-dominated cosmology, co-annihilation with $N_1$ allows scalar DM in the range $300~\mathrm{GeV}\lesssim M_{A^0}\lesssim500~\mathrm{GeV}$ to reproduce the observed relic abundance, whereas this region remains underabundant in the corresponding IHDM scenario. In this mass range, generally we have small mass splittings between $A^0$ and dark sector particles $H^\pm$ and $N_1$. Such near-degeneracy, together with sufficiently small Yukawa couplings, can lead to macroscopic decay lengths. The resulting lifetime, however, depends on both the available phase space and the relevant Yukawa couplings. Two representative mass hierarchies can be distinguished:
$\text{(i)}\,\, M_{N_1}>M_{H^\pm}>M_{A^0},\,\,\,
\text{(ii)}\,\, M_{H^\pm}>M_{N_1}>M_{A^0}$.
The position of $H^0$ in the spectrum determines whether additional neutral-scalar decay channels are kinematically accessible but does not affect this classification.

In the first hierarchy, $N_k$ can decay through the visible channel $N_k\to\ell_\alpha^\pm H^\mp$ and through the neutral channels $N_k\to\nu_\alpha H^0$ and $N_k\to\nu_\alpha A^0$, whenever they are kinematically allowed. The corresponding partial widths are
\begin{align}
&\Gamma_{N_k\rightarrow\ell_\alpha^+H^-}
=\Gamma_{N_k\rightarrow\ell_\alpha^-H^+}
=
\frac{|Y^\nu_{k\alpha}|^2
\left(M_{N_k}^2+m_{\ell_\alpha}^2-M_{H^\pm}^2\right)}
{32\pi M_{N_k}^3}
\sqrt{
\left(M_{N_k}^2-m_{\ell_\alpha}^2-M_{H^\pm}^2\right)^2
-4m_{\ell_\alpha}^2M_{H^\pm}^2},\\
&\Gamma_{N_k\rightarrow\nu H^0/A^0}
=
\sum_\alpha
\frac{|Y^\nu_{k\alpha}|^2}{32\pi M_{N_k}^3}
\left(M_{N_k}^2-M_{H^0/A^0}^2\right)^2.
\label{eq:N1decays}
\end{align}
For small $\Delta m_{N_1}$ and sufficiently suppressed Yukawa couplings, $N_1$ can become long-lived. Since $N_1$ is electrically neutral, it does not itself leave a charged track; instead, its visible decay modes can produce displaced vertices containing charged leptons. At lepton colliders, $N_1$ pairs can be produced through $t$-channel $H^\pm$ exchange. However, the cross-section is strongly suppressed in the small-Yukawa regime required for a long-lived $N_1$, making direct observation challenging.

In the second hierarchy, the two-body decay $H^\pm\to N_k\ell_\alpha^\pm$ is kinematically allowed, with partial width
\begin{align}
\Gamma_{H^\pm\rightarrow N_k\ell_\alpha^\pm}
=
\frac{|Y^\nu_{k\alpha}|^2
\left[M_{H^\pm}^2-(M_{N_k}+m_{\ell_\alpha})^2\right]}
{8\pi M_{H^\pm}}
\times
\sqrt{1-\left(\frac{M_{N_k}-m_{\ell_\alpha}}{M_{H^\pm}}\right)^2}
\sqrt{1-\left(\frac{M_{N_k}+m_{\ell_\alpha}}{M_{H^\pm}}\right)^2}.
\label{eq:decayetalepton}
\end{align}
The charged scalar can also undergo the gauge-mediated decays $H^\pm\to A^0W^{\pm}$ when kinematically allowed. Because the scalar spectrum relevant to the low-mass region is highly compressed, the $W$ boson is generally off-shell.
An especially interesting possibility arises when
$\Delta m_{H^\pm}=M_{H^\pm}-M_{A^0}
\sim\mathcal{O}(100)~\mathrm{MeV}$
and lies above the pion threshold. In this regime, $H^\pm\to A^0\pi^\pm$ can dominate if the Yukawa coupling at the $H^\pm N_1\ell$ vertex is sufficiently small to suppress the competing leptonic decay. Its partial width is
\begin{align}
\Gamma_{H^\pm\rightarrow A^0\pi^\pm}
=
\frac{f_\pi^2g^4}{m_W^4}
\frac{\left(M_{H^\pm}^2-M_{A^0}^2\right)^2}
{512\pi M_{H^\pm}}
\times
\sqrt{1-\left(\frac{M_{A^0}-m_\pi}{M_{H^\pm}}\right)^2}
\sqrt{1-\left(\frac{M_{A^0}+m_\pi}{M_{H^\pm}}\right)^2},
\label{eq:decayetapion}
\end{align}
where $f_\pi$ is the pion decay constant and $m_\pi$ is the pion mass. The limited phase space suppresses the decay width and can make $H^\pm$ sufficiently long-lived to travel a measurable distance inside the detector. Moreover, the pion produced in this decay is typically too soft to be reconstructed. The charged scalar therefore leaves a track in the inner detector that terminates after it decays, giving rise to the characteristic disappearing-track signature~\cite{ATLAS:2017oal}.

Unlike direct $N_1$ pair production, $H^+H^-$ production receives gauge-mediated contributions and is not suppressed by small Yukawa couplings. As shown in Fig.~\ref{fig:XS}, its production cross-section can remain sizable for charged-scalar masses below approximately $500$ GeV. This makes the compressed scalar region particularly interesting for dedicated disappearing-track searches at the HL-LHC and future lepton colliders.

\subsection{Scotogenic model: Kination-dominated cosmology}

An early kination-dominated epoch significantly enlarges the cosmologically viable parameter space. The enhanced expansion rate increases the relic abundance of parameter points that are underabundant in the standard cosmology, opening the scalar DM mass range $200~\mathrm{GeV}\lesssim M_{A^0}\lesssim1~\mathrm{TeV}$. In particular, Fig.~\ref{fig:coanni} shows that, for sufficiently low reheating temperatures, viable points occur for $200~\mathrm{GeV}\lesssim M_{A^0}\lesssim300~\mathrm{GeV}$.
Although these points generally retain a compressed scalar spectrum, with $\Delta m_{H^\pm}\lesssim1$ GeV, the fermion mass splitting $\Delta m_{N_1}$ can extend over essentially the full scanned range, up to $\mathcal{O}(10^4)$ GeV. Thus, co-annihilation with $N_1$ is not essential in this region; the required relic density enhancement is instead predominantly supplied by the modified cosmological expansion.

Kination does not introduce new collider production mechanisms, but it changes the masses and hierarchies that are compatible with the observed relic abundance. The compressed subset of the parameter space can still lead to long-lived $N_1$ or $H^\pm$ signatures analogous to those discussed in the radiation-dominated case. The extension to $M_{A^0}<300$ GeV is particularly interesting because the associated inert scalars are lighter and can consequently have larger electroweak production cross-sections. Conversely, when $\Delta m_{N_1}$ is large, the singlet fermion effectively decouples, and the collider phenomenology becomes more similar to that of the IHDM, although in a lower scalar mass region made viable by kination.

For larger Yukawa couplings and sufficient phase space, the decay $H^\pm\to N_k\ell_\alpha^\pm$ can be prompt. Charged-scalar pair production can then lead to $\ell_\alpha^\pm\ell_\beta^\mp+\slashed{E}_T$
final states, where the missing energy is carried by the neutral $Z_2$-odd particles. At lepton colliders, direct $N_1$ pair production through $t$-channel $H^\pm$ exchange can also become relevant for larger Yukawa couplings, provided that $N_1$ is kinematically accessible. Depending on its lifetime and available decay modes, this may produce prompt or displaced leptons accompanied by missing energy.

Thus, compared with both the IHDM and the radiation-dominated scotogenic scenario, the kination-dominated framework provides a broader cosmologically viable mass range and a more varied dark sector spectrum. This leads to complementary collider signatures involving missing transverse energy, prompt leptons, displaced vertices, and disappearing charged tracks. A quantitative assessment of their experimental reach would require a dedicated detector-level analysis.
\vspace{-0.5cm}
\section{Conclusion}
In this section, we summarize our main findings on scalar DM in the scotogenic model and highlight the role of an early kination-dominated epoch in shaping the viable parameter space.

\par In the original Scotogenic model, the scalar dark matter in the $(300-500)$ GeV region can reproduce the observed relic abundance in the standard radiation-dominated cosmology when the lightest singlet fermion is nearly degenerate with the dark matter. The resulting compressed spectrum can lead to a long-lived $N_1$ or charged scalar, with disappearing charged tracks providing the most promising collider signature. Our analysis substantially extends this picture by incorporating an early kination-dominated epoch. The enhanced expansion rate causes earlier freeze-out and raises the relic abundance of parameter points that remain underabundant even after including fermionic coannihilation. This opens a broader viable mass range and relaxes the need for an extremely compressed dark sector spectrum. Furthermore, we determine the reheating temperature required for each viable point and examine the resulting parameter space using indirect detection, CMB, direct detection and collider probes. We find that the $(100-200)$ GeV region is strongly constrained by existing indirect searches, whereas much of the surviving $200$ GeV - 1 TeV region can be tested by CTA, next-generation direct detection experiments, the HL-LHC and future lepton colliders.
\vspace{-0.5cm}
\acknowledgments
\noindent RK is supported by the National Research Foundation of Korea under Grant NRF-2023R1A2C100609111. The work of SM is supported by KIAS Individual Grants (PG086002) at Korea Institute for Advanced Study. The work of AR is supported by Basic Science Research Program through the National Research Foundation of Korea (NRF) funded by the Ministry of Education through the Center for Quantum Spacetime (CQUeST) of Sogang University (RS-2020-NR049598) and by the Ministry of Science and ICT with grant number RS-2025-24523022.
\bibliographystyle{utphys}
\bibliography{bibliography.bib}
\end{document}